\documentclass{pasj00}
\draft

\begin{document}
\SetRunningHead{Terada et al. in page-head}
{Anisotropic Resonance Scattering in MCV hot plasmas}
\Received{2003/07/22}
\Accepted{yyyy/mm/dd}

\title{The Anisotropic Transfer of Resonance Photons \\
in Hot Plasmas on Magnetized White Dwarfs}
\author{Yukikatsu \textsc{Terada}\altaffilmark{1},
Manabu \textsc{Ishida}\altaffilmark{2}, 
Kazuo \textsc{Makishima}\altaffilmark{1,3}}
\altaffiltext{1}{Cosmic Radiation Laboratory, RIKEN, \\
2-1 Hirosawa, Wako, Saitama, 351-0198 Japan}
\email{terada@riken.go.jp}
\altaffiltext{2}{Department of Physics, Tokyo Metropolitan University, \\
1-1 Minami-Ohsawa, Hachioji-shi, Tokyo, 192-0397 Japan}
\email{ishida@phys.metro-u.ac.jp}
\altaffiltext{3}{Department of Physics, Science, The University of Tokyo,\\
7-3-1, Hongo, Bunkyo-ku, Tokyo, 113-0033 Japan}
\email{maxima@phys.s.u-tokyo.ac.jp}


\maketitle
\begin{abstract}
In order to confirm the anisotropic effect of resonance photons
in hot accretion columns on white dwarfs in magnetic cataclysmic variables,
proposed by Terada \etal (2001), systematic studies with {\it ASCA} of 
7 polars and 12 intermediate polars are performed.
The equivalent widths of He-like Fe K$_\alpha$ line of polars
are found to be systematically modulated at their spin periods
in such a way that it increases at the pole-on phase.
This implies that the anisotropic mechanism is commonly operating among polars.
On the other hand, those of intermediate polars are statistically consistent 
with being unmodulated with an upper limit of 1.5 times modulation.
This may be due to a different accretion manner like aurora curtain
\citep{Rosen_88_Curtain_Model} 
so that the plasma becomes optically thin along the horizontal axis
also for the resonance lines, or 
because of larger optical depths for Compton scattering
if the emission regions have the same coin-like shapes as polars. 
\end{abstract}

\section{Introduction}
\label{section:introduction}
A magnetic cataclysmic variable (MCV) is a close binary consisting 
of a white dwarf (WD) and a mass-donating companion star.
Matter that spills over the Roche lobe of the companion star accretes
on a magnetic pole of the WD, forming a post-shock hot plasma 
(Patterson 1994 and references therein).
According to the systematic X-ray study of MCVs by {\it Ginga},
their X-ray spectra can be described by 
optically thin thermal bremsstrahlung continua with 
Fe lines \citep{Ishida_91_PhD}.
These X-ray characteristics are approximately the same 
for both subgroups of MCVs, polars and intermediate polars (IPs).
Plasma diagnostics with atomic lines, performed with {\it ASCA}
(\cite{Fujimoto_97_asca_exhya}; \cite{Ezuka_99}), 
confirmed the multi-temperature structure of the plasma as
had been predicted by \citet{Hoshi_73} and \citet{aizu}.

Considering the typical size ($\sim 10^{7}$ cm) and
density ($\sim 10^{15 \mbox{--} 16}$ cm$^{-3}$) 
of an accretion column of an MCV, 
the plasma is not completely optically thin;
it is estimated to be marginally optically thick
for Compton (or Thomson) scattering with an optical depth of order 0.1.
Under such conditions, 
the atomic line structure is expected to be suppressed by 10\%--20\%,
because the line photons would experience large shifts reaching
$\sim$ 1 keV as they Compton scatter off hot electrons.
At the same time, the plasma is optically thick 
for resonant photons of atomic lines 
from abundant heavy elements, including Fe in particular
\citep{Fujimoto_97_asca_exhya}.
Then, the line structure of resonance transitions would be more suppressed,
because the resonant photons would be effectively trapped in the column,
and escape most easily when they Compton scatter out of the resonance.
Nevertheless, we usually observe many ionized atomic lines 
in the X-ray spectra of MCVs \citep{Ezuka_99}, and 
apparent metal abundances derived from the equivalent width (EW)
of these lines are not much lower than one times solar.
In addition, some MCVs called Pole-on line emitters (POLES; Paper I),
including AX~J2315$-$0592 \citep{Misaki_96_axj2315}, 
RX~J1802.1+1804 \citep{Ishida_98_rxj1802_asca}, and 
AX~J1842.8$-$0423 \citep{Terada_1999_paper1}, 
exhibit extremely strong He-like Fe-K lines, having an EW of $\sim 4000$ eV, 
implying apparent Fe abundances of $\sim$ three times solar.

In order to solve the puzzle of extremely strong Fe lines,
we have developed a scenario of Fe-K line collimation along the vertical
axis of the accretion column incorporating resonance scattering 
(\cite{Terada_1999_paper1}; \cite{Terada_2001_paper2}, hereafter Paper I;
\cite{Terada_2002_phD}).
The anisotropy occurs when the accretion column has a flat shape,
and is augmented by the strong longitudinal velocity gradient in the 
accretion flow, which invalidates the resonance condition for line photons
when they propagate along the vertical axis.
Since the Fe ions are heavy,
their thermal resonance width (typically $\sim3$ eV for a 10 keV plasma)
is easily exceeded by the vertical Doppler shift 
($\sim10$ eV from the bottom to the top of the column).
Therefore, the resonant trapping of Fe-K photons 
is preferentially reduced along the vertical axis, 
and hence the resonance line photons
are collimated along the vertical direction.
This scenario can explain both the lack of line destruction 
and the enigmatic POLE phenomenon.

We have numerically confirmed the anisotropic propagation of resonance photons
via extensive Monte Carlo simulations (Paper I),
employing the analytic solution to the post-shock flow by \citet{aizu}.
The results show that the directional emissivity of the resonant Fe-K line
photons is indeed enhanced up to a factor of $\sim2\mbox{--}2.5$.
In addition, we have confirmed, 
through {\it ASCA} observations of the polar V834 Centauri,
that the EW of its He-like Fe K$_\alpha$ line
increases during pole-on phases of its rotation (Paper I).
The proposed mechanism has thus been confirmed 
through calculation, and observation of a few particular polars.

According to our Monte Carlo simulations,
the anisotropic propagation of resonance Fe-line photons 
should occur for a rather wide range of plasma parameters 
in the accretion column (Paper I).
We hence expect to observe the rotational modulation of the resonant 
Fe-K line EW, not only in the particular case of V834 Centauri, 
but also of MCVs in general.
In order to examine this conjecture, here we systematically analyze
the {\it ASCA} data of 7 polars and 12 IPs.

\section{Observation}
\label{section:observation}
The central aim of this paper is to search the data for the
possible dependence of the Fe line EW on the angle, $\theta$,
between our line-of-sight and the axis of the accretion column.
We therefore need to perform spin-phase-resolved spectroscopy
of the target MCVs. Accordingly, we have analyzed all the 9 polars and 13 IPs 
observed with {\it ASCA} satellite \citep{Tanaka_94_asca94}.
After discarding objects which are too X-ray faint ($<$ 0.05 GIS cnt s$^{-1}$)
and those with insufficient exposure ($<$ 20 ksec), we are left with 
7 polars and 12 IPs, as listed in table \ref{tbl:target_table},
including V834 Centauri, results for which are reported in Paper I.
Some objects (V834 Cen, EK UMa, and RX~J1015+0904) are the approved
targets of our proposal for this purpose, and the others are archived
or published observations.
Their geometries, i.e., the inclination angle $i$ and the pole colatitude
$\beta$, are obtained by polarimetric observations 
in the optical to ultraviolet band, and summarized in table 4 of Paper I.

For each target, we accumulated the GIS 
\citep{Ohashi_96_gis1,Makishima_96_gis2} and SIS \citep{Burke_91_sis3} events
within a circle of radius $4'\hspace{-3pt}.5$ centered on the object,
employing the following data-selection criteria.
We discarded the data during {\it ASCA} passage 
through the South Atlantic Anomaly, and when the field of view of {\it ASCA}
was within 5$^\circ$ of the Earth's rim.
Furthermore, we discarded the GIS data during occasional errors
in the on-board CPU, as well as the SIS data acquired 
when the field of view was within 10$^\circ$ of the bright Earth rim
or when the spacecraft crossed the day-night transition zones.

\section{Data Analysis and results}
\label{section:analysis_and_results}
\subsection{X-ray light curves}
\label{subsection:ana_lc}
\begin{figure*}[bht]
\begin{center}
\FigureFile(15.0cm,6cm){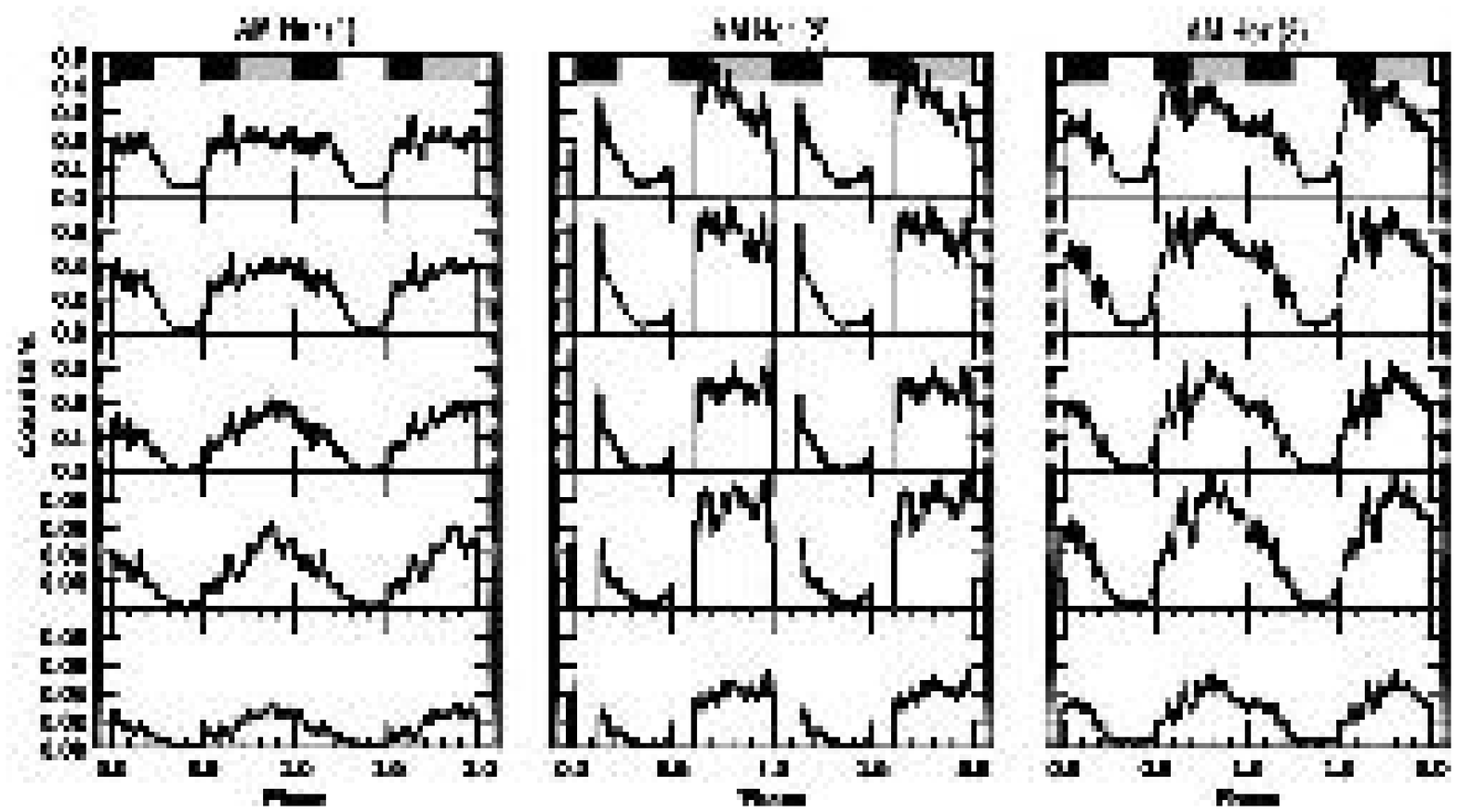}\\
\FigureFile(5.0cm,6cm){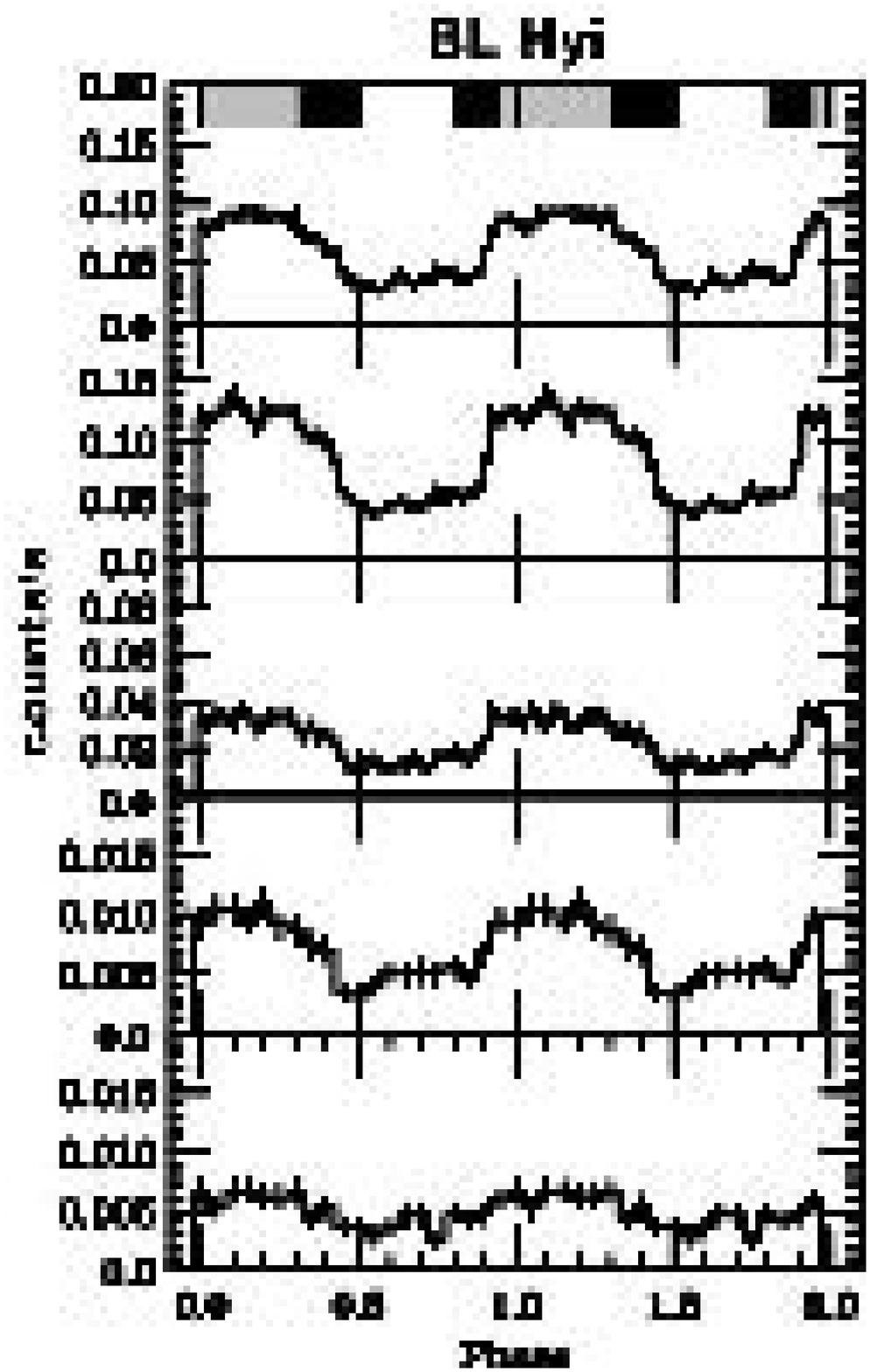}
\FigureFile(5.0cm,6cm){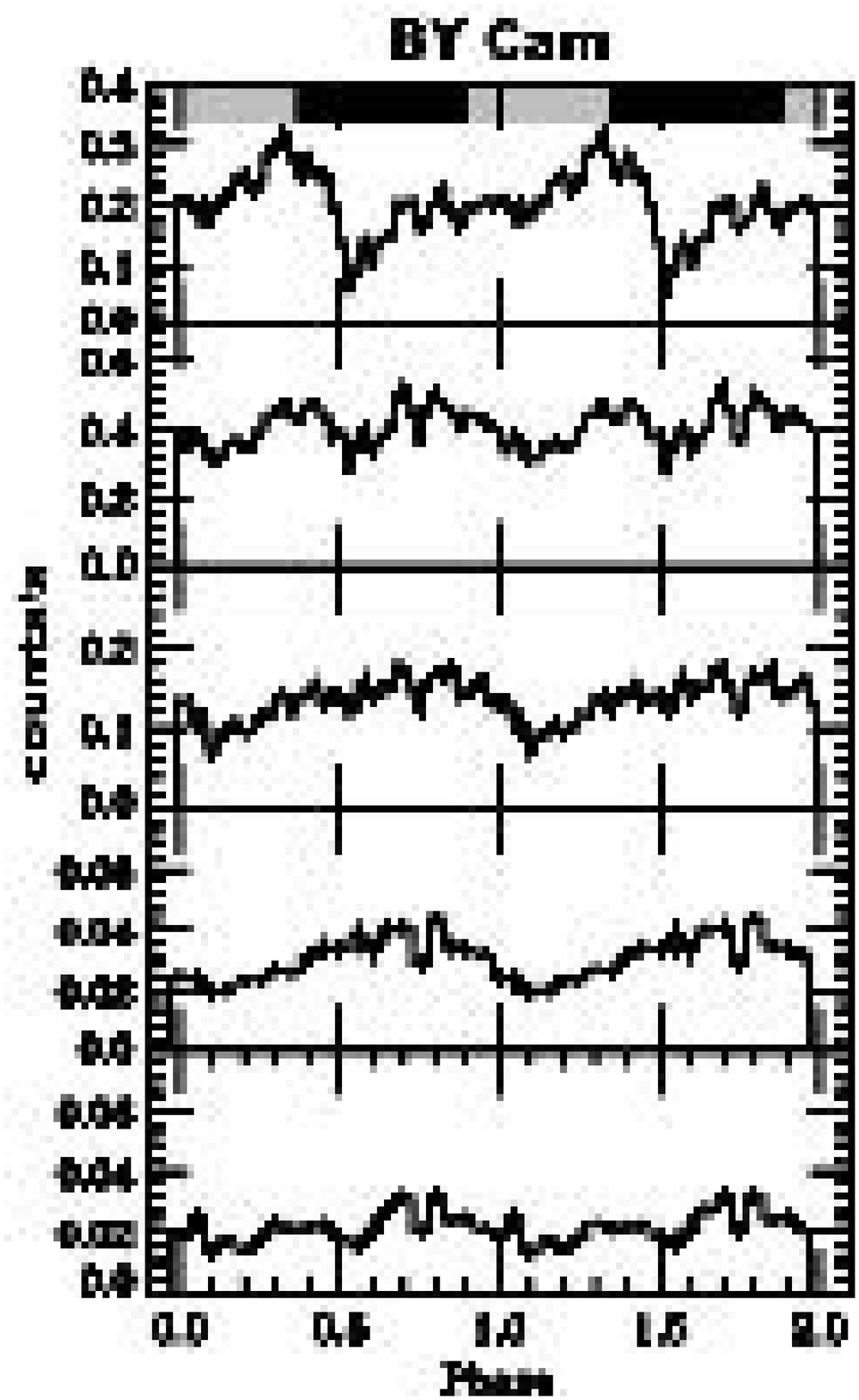}
\FigureFile(5.0cm,6cm){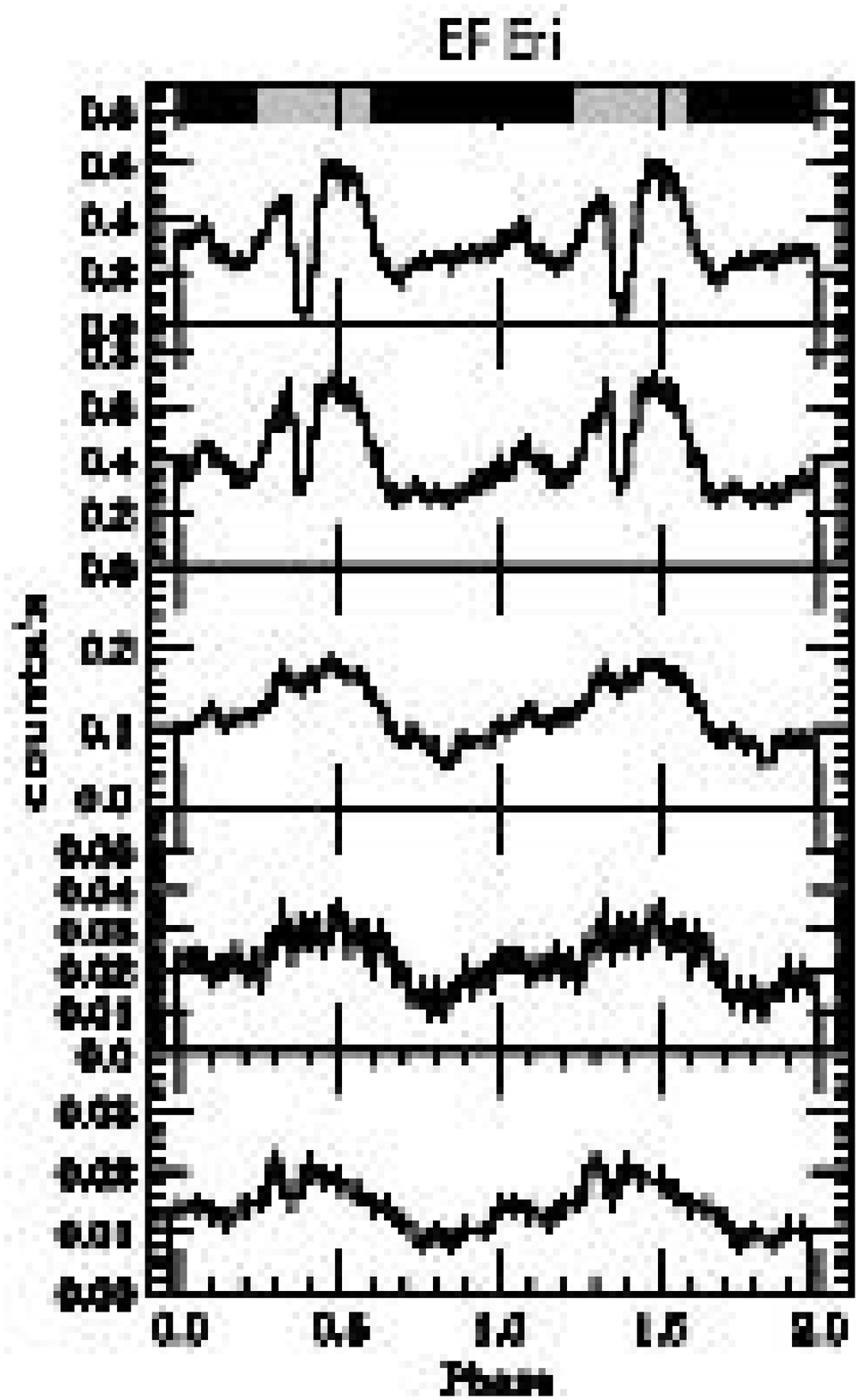}
\end{center}
\caption{X-ray light curves of the polars observed with {\it ASCA},
folded on their spin periods listed in table \ref{tbl:target_table}.
The respective phase 0.0 is given in table \ref{tbl:target_table}.
For each object, light curves in five energy bands 
(0.5--1.5, 1.5--4.0, 4.0--6.2, 6.2--7.2, 7.2--10.0 keV from top to bottom) 
are presented. The pole-on and side-on phases are shown at the top of each
panel by the gray and black bands, respectively (see the text). 
The data of V834 Cen is reported in Paper I.}
\label{fig:other_polars_lc}
\end{figure*}

To perform the phase resolved analysis, we folded X-ray light curves 
of the selected objects in table \ref{tbl:target_table}
on their rotational periods.
As shown in figure \ref{fig:other_polars_lc},
the polars mostly exhibit single-peaked folded light curves 
in the high-energy band, indicating that we are observing a single pole.
The deep X-ray minima seen in some polars to higher energies
are due to the self-eclipse of the emission region, while
the residual mild X-ray intensity modulations 
during the uneclipsed phase are explained in terms of the
varying contribution from the scattered and/or reflected component
from the surface of the white dwarfs 
\citep{Beardmore_95_polar_modulation_reflection}.
Therefore, the phase of the maximum X-ray intensities roughly corresponds to 
the pole-on phase ($\theta\sim0$), 
or the phase where our line of sights comes closest
to the accretion-column axis (minimum $\theta$).
In addition, in figure \ref{fig:other_polars_lc},
the light curves often exhibit absorption dips in softer energies 
at the X-ray maximum phase.
Since these dips arise due to photoelectric absorption by 
pre-shock matter in the accretion column, 
they can be used as an additional indicator of the pole-on phase.
In this way, we have determined the pole-on and side-on phases 
of our target sources, as shown 
by the gray and black bands, respectively, in figure \ref{fig:other_polars_lc}.
Utilizing the geometric parameters summarized in table 4 of Paper I,
we have excluded the phase of $\theta > 90^\circ$,
in which the emission region is eclipsed by the white dwarfs.

For the phase determination of the IPs 
shown in figure \ref{fig:other_intermediate_polars_lc1}, 
we have to rely solely on the X-ray spin modulations,
because of the lack of optical polarization.
According to the phase-resolved spectroscopy with {\it Ginga} 
\citep{Ishida_91_PhD},
the X-ray minimum phase is thought to correspond to the pole-on phase,
because the X-ray modulation is due to photoelectric absorption 
by the pre-shock matter \citep{Rosen_88_Curtain_Model}, 
as evidenced by the deeper X-ray modulations in softer energy bands.
This idea of self-absorption is supported by Doppler measurements
of He II emission lines from the pre-shock matter 
(\cite{Hellier_87_doppler_HII_1}, \yearcite{Hellier_90_doppler_HII_2}).
We have thus defined the pole-on and side-on phases as 
shown by the gray and black bands, respectively, in figure
\ref{fig:other_intermediate_polars_lc1}.

\subsection{Phase-resolved spectra}
\label{subsection:ana_spec}
We have accumulated the {\it ASCA} spectra separately over the pole-on
and side-on phases, as shown in figures \ref{fig:polars_spec}
and \ref{fig:intermediate_polars_spec} for the polars and intermediate
polars, respectively.
To quantify the EWs of the Fe lines,
we adopted the standard spectral model, i.e., 
a photoelectrically absorbed bremsstrahlung continuum
with three narrow Gaussians (\cite{Ezuka_99}; Paper I),
the latter representing neutral (at 6.4 keV), He-like (at $\sim6.7$ keV),
and H-like (at $\sim6.9$ keV) Fe-K$_\alpha$ lines.
The three lines can be resolved with the SIS, but not with the GIS.
Accordingly, we fixed the centroid energies of the first Gaussian
at 6.4 keV, fixed the ratio of the centroid energies of the others
at the theoretical value of 1.042, and assumed them to be narrow.
The energy range for the continuum fitting was optimized in the manner
described in section 3.2 of \citet{Ezuka_99}.
The SIS and GIS spectra were fitted simultaneously.
The fits are successful, as shown in 
figures \ref{fig:polars_spec} and \ref{fig:intermediate_polars_spec},
where we compare predictions of the best-fit model with the actual data.

Figure \ref{fig:polars_ew} compares the EWs of the H-like and 
He-like Fe-K$_\alpha$ lines of individual polars for their pole-on and 
side-on phases. Figure \ref{fig:intermediate_polars_ew} presents the same 
comparison for the IPs.
For reference, the EWs of the fluorescent Fe-K$_\alpha$ line,
which is not emitted from the hot accretion columns but 
probably from the WD surfaces, are also plotted in the same figures.
As is clearly demonstrated by figure \ref{fig:polars_ew},
the EWs of the He-like Fe line of our sample polars are
systematically larger in the pole-on phase than in the side-on phase,
although the enhancement in individual objects is insignificant,
and that of the H-like line is consistent with being unmodulated.
Quantitatively, the enhancement of the H-like and He-like Fe K lines,
weighted by the absolute values of their EWs, are
$\zeta^{\rm P}_{\rm OBS}({\rm H}) = 1.05 \pm 0.47$ and
$\zeta^{\rm P}_{\rm OBS}({\rm He}) = 1.88 \pm 0.72$, respectively
(90\% confidence errors).
The significance is slightly reduced to
$\zeta^{\rm P}_{\rm OBS}({\rm H}) = 1.05 \pm 0.53$ and
$\zeta^{\rm P}_{\rm OBS}({\rm He}) = 1.89 \pm 0.86$,
when we exclude the data of V834 Centauri, which
were already presented in Paper I.
Our result for the sample of polars therefore confirms 
the enhancement of the He-like line in a statistical sense.
For the IPs, we see no modulation in either Fe line
(figure\ref{fig:intermediate_polars_ew});
the enhancement of the H-like and He-like Fe K lines are
$\zeta^{\rm IP}_{\rm OBS}({\rm H}) = 1.02 \pm 0.47$ and
$\zeta^{\rm IP}_{\rm OBS}({\rm He}) = 1.14 \pm 0.36$, respectively.

\section{Discussion}
\label{section:discussion}
In order to confirm the general validity of 
the anisotropic transfer scenario of resonance photons in MCVs
pointed out in Paper I, 
we have systematically analyzed the {\it ASCA} data of 
7 polars and 12 intermediate polars.
Through phase-resolved analysis of the ionized Fe K$_\alpha$ lines,
the EWs of the He-like Fe lines of the polars have been found to be 
systematically enhanced in their pole-on phases. 
Although the significance is only 90\%,
and the individual cases were insignificant except for V834 Centauri (Paper-I),
the statistical trend is in support of our scenario.
To better visualize this result, 
we plotted in figure \ref{fig:polar_theta_xi} left panel
the EWs of the He-like Fe lines of the polars relative 
to the phase averaged value, as a function of $\theta$.
There,we also plot the results of our Monte Carlo simulations based on 
representative parameters (see caption).
Thus, the observed spin-phase dependence of the Fe-K line EWs 
is consistent with the calculation, although within rather large errors.

The observed He-like Fe line is in fact a blend of resonance, 
forbidden, and intercombitation lines, where 
the latter two do not suffer the resonance effect.
Since the H-like line is a pure resonance line,
we may expect to observe a larger spin modulation 
for this line compared to that of the He-like blend.
According to our Monte Carlo simulations (Paper I; \cite{Terada_2002_phD}),
however, the expected enhancement of the H-like Fe line is
actually smaller than that of the He-like resonance line,
because the former photons are preferentially produced in a higher
portion of the post-shock column, which
has lower densities and smaller velocity differences 
than those at the bottom of the column \citep{aizu}.
As presented in figure \ref{fig:sim_rslt_he_h_temp},
we expect almost the same enhancement of H-like Fe line
as that of the He-like blend.
Furthermore, the H-like Fe-K line is subject to 
rather large measurement errors, due to its relatively poor statistics and 
its possible confusion with the Fe K$_\beta$ emission line from He-like iron.
As a result, the measurements of the H-like Fe line EWs
(figures \ref{fig:polars_ew}, \ref{fig:intermediate_polars_ew}, and
\ref{fig:polar_theta_xi} right) are consistent with the prediction.

As for IPs (figure \ref{fig:intermediate_polars_ew}),
the modulations of both line species have 
been found to be statistically insignificant.
One possible explanation is that the accretion streams of IPs have
a curtain-like shape \citep{Rosen_88_Curtain_Model}, with a much-reduced
optical depth in the lateral direction,
so that the anisotropic resonance effect is suppressed.
Alternatively, the emission region may have a column shape like polars,
and the lack of spin modulation in the Fe line EWs 
may be attributed to higher electron densities $n_e$ in the plasma,
as is indicated by their larger volume emission measure 
(VEM, described as $\int n_e^2 dV$, where $V$ is the plasma volume)
than those of polars.
Then, the enhanced Compton scattering by hot electrons reduces 
the anisotropic effect on the resonance photons (see figure 8 in Paper I).
In fact, by comparing figure \ref{fig:polars_ew} with 
figure \ref{fig:intermediate_polars_ew}, we find that the IPs exhibit 
systematically smaller EWs of Fe K lines than the polars.

In order to examine the latter possibility for the IPs,
we plot in figure \ref{fig:mcv_emission_measure} 
VEMs of the polars and IPs listed in table \ref{tbl:target_table},
which shows that the VEMs of IPs are one-to-two orders of magnitude larger 
than those of polars. 
Then, adopting the same discussion as in section 5.2 of Paper I, 
we estimate that the accretion column of an IP has
an order-of-magnitude higher electron density of 
$n_{\rm e}\sim10^{17}$ cm$^{-3}$,
an order-of-magnitude lower column height of $h \sim 10^6$ cm, and 
several times larger column radius of $r \sim 10^7$ cm,
all compared with polars.
Under such conditions, the optical depth to Compton scattering is 
$\sim 1.0$, whereas it is only 0.4 for polars.
Therefore, the interpretation may actually work.

The Fe-K line photons provide the best diagnostics of the resonance effects,
because Fe ions are massive enough for the thermal Doppler effect to fall
significantly below the bulk-motion Doppler shift.
Although the {\it ASCA} data of MCVs are rather limited
both in statistics for phase-resolved analyses and energy resolution,
a quantum jump in this research subject may be realized 
by the {\it ASTRO-E 2} satellite to be launched in 2005,
with its energy resolution reaching $\sim$ 10 eV
around the Fe-K$_\alpha$-line range for the first time.

\bigskip
Finally, we thank the members of the {\it ASCA} team 
for spacecraft operation and data acquisition.


\clearpage
\begin{table*}[hbt]
\begin{center}
\caption{Objects analyzed in the present paper.}
\label{tbl:target_table}
\begin{tabular}{lp{0.5cm}lllllllc}
\hline 
{object}&	
\multicolumn{2}{c}{Date of observation}&\multicolumn{2}{c}{Exp.\ (ksec)\footnotemark[$*$]}&
\multicolumn{2}{c}{Cnt Rate}&
\multicolumn{1}{c}{phase 0.0 (HJD)}&
\multicolumn{1}{c}{Spin period}&{ref.$^\dagger$}\\
{name}&{}&\multicolumn{1}{c}{(UT)}&	{SIS}&	{GIS}&	{SIS}&	{GIS}&{
in figures \ref{fig:other_polars_lc} and \ref{fig:other_intermediate_polars_lc1}}&{(days)}&{}\\
\hline 
\multicolumn{4}{l}{\bf Polars}\\
{V834 Cen}&	{(1)}&{1994/03/03.64--04.13}&	{21.7}&		{23.2}&		{0.26}&		{0.18}&
		{2445048.9500}&	{0.070498}&{1}\\
{}&		{(2)}&{1999/02/09.93--11.72}&	{62.3}&		{53.5}&		{0.25}&		{0.19}&
		{}&{}&{}\\
{AM Her}&	{(1)}&{1993/09/27.22--28.27}&	{30.9}&		{43.5}&		{0.61}&		{0.43}&
		{2443014.76614}&	{0.128927}&{2}\\
{}&		{(2)}&{1995/03/06.84--07.22}&	{18.3}&		{16.8}&		{0.92}&		{0.66}&
		{}&{}&{}\\
{}&		{(3)}&{1995/03/09.03--10.03}&	{45.5}&		{41.1}&		{0.89}&		{0.65}&	
		{}&{}&{}\\
{BL Hyi}&	{}&{1994/10/11.47--12.59}&	{41.4}&		{43.5}&		{0.21}&		{0.16}&	
		{2450379.4725}&	{0.078915}&{3}\\
{BY Cam}&	{}&{1994/03/11.49--12.16}&	{29.5}&		{34.7}&		{0.98}&		{0.68}&
		{2446138.8202}&	{0.13979}&{4}\\
{EF Eri}&	{}&{1993/07/23.90--24.77}&	{35.9}&		{39.0}&		{0.93}&		{0.61}&
		{2443894.6824}&	{0.056266}&{5}\\
{\small RXJ1015+0904}&	{}&{1999/05/04.08--05.41}&	{54.0}&		{54.0}&		{0.09}&		{0.07}&
		{2451302.500}&	{0.055471}&{6}\\
{V2301 Oph}&	{}&{1998/09/28.49--30.21}&	{70.6}&		{73.0}&		{0.21}&		{0.15}&
		{2448071.02014}&	{0.078450}&{7}\\
\hline 
\multicolumn{4}{l}{\bf Intermediate Polars}\\
{AO Psc}&	{}&{1994/06/22.29--24.56}&	{81.0}&		{84.3}&		{0.57}&		{0.48}&
		{2444883.92074}&	{0.00931948}&{8}\\
{BG CMi}&	{(1)}&{1996/04/14.76--15.97}&	{42.0}&		{43.4}&		{0.16}&		{0.17}&
		{2450186.5}&	{0.0105729}&{9}\\
{}&		{(2)}&{1996/04/17.44--18.53}&	{40.9}&		{41.5}&		{0.15}&		{0.16}&
		{}&{}&{}\\
{EX Hya}&	{}&{1993/07/16.45--17.64}&	{36.0}&		{38.8}&		{2.79}&		{1.50}&
		{2437699.8914}&	{0.0465465}&{10}\\
{FO Aqr}&	{}&{1993/05/20.93--02.04}&	{37.8}&		{38.0}&		{0.27}&		{0.32}&
		{2446097.243668}&	{0.01451911}&{11}\\
{PQ Gem}&	{(1)}&{1994/11/04.91--07.10}&	{76.2}&		{80.2}&		{0.37}&		{0.30}&
		{2449297.9730}&	{0.009645994}&{12}\\
{}&		{(2)}&{1999/10/19.71--20.31}&	{35.0}&		{42.9}&		{0.37}&		{0.28}&
		{}&{}&{}\\
{\small RXJ1712-2414}&	{}&{1996/03/18.85--21.24}&	{81.4}&		{84.2}&		{0.79}&		{0.65}&
		{2450159.5}&	{0.010737}&{13}\\
{TV Col}&	{}&{1995/02/28.25--01.44}&	{36.7}&		{39.9}&		{0.65}&		{0.54}&
		{2447139.524}&	{0.02211}&{14}\\
{TX Col}&	{}&{1994/10/03.29--04.42}&	{39.9}&		{45.1}&		{0.22}&		{0.16}&
		{2449627.5}&	{0.02212}&{6}\\
{V1062 Tau}&	{}&{1998/02/16.89--18.48}&	{57.5}&		{14.7}&		{0.23}&		{0.24}&
		{2450859.5}&	{0.04313}&{9}\\
{V1223 Sgr}&	{}&{1994/04/24.18--25.96}&	{57.0}&		{59.2}&		{1.23}&		{1.00}&
		{2445626.13067}&	{0.00862854}&	{17}\\
{V405 Aur}&	{(1)}&{1996/10/05.56--07.47}&	{47.5}&		{41.1}&		{0.34}&		{0.23}&
		{2449689.573466}&	{0.006313154}&{15}\\
{}&		{(2)}&{1999/03/21.40--22.44}&	{39.0}&		{43.3}&		{0.38}&		{0.26}&
		{}&{}&{}\\
{XY Ari}&	{(1)}&{1995/08/07.00--07.91}&	{34.6}&		{34.9}&		{0.14}&		{0.15}&	
		{2449935.5}&	{0.00238773}&{16}\\
{}&		{(2)}&{1996/01/28.21--29.73}&	{58.5}&		{61.0}&		{0.16}&		{0.16}&
		{}&{}&{}\\
{}&		{(3)}&{1996/02/18.97--19.56}&	{24.4}&		{25.6}&		{0.13}&		{0.14}&
		{}&{}&{}\\
\hline 
\multicolumn{10}{@{}l@{}}{\hbox to 0pt{\parbox{180mm}{\footnotesize
\par\noindent
References; 1 \citet{V834_Spin}, 2 \citet{AMHer_Spin}, 
3 \citet{BLHyi_Spin}, 4 \citet{BYCam_Spin}, 5 \citet{EFEri_Spin}, 
6 \citet{TXCol_Spin}, 7 \citet{V2301_Spin}, 8 \citet{AOPsc_Spin},
9 \citet{polar_review}, 10 \citet{EXHya_spin}, 11 \citet{FOAqr_Spin},
12 \citet{PQGem_Spin}, 13 \citet{RXJ1712_Spin}, 14 \citet{TVCol_Spin},
15 \citet{V405_Spin}, 16 \citet{XYAri_Spin}, and 17 {\it ASCA} Timing Analysis.
}\hss}}
\end{tabular}
\end{center}
\end{table*}

\clearpage

\begin{figure*}[hbt]
\begin{center}
\FigureFile(5.0cm,6.0cm){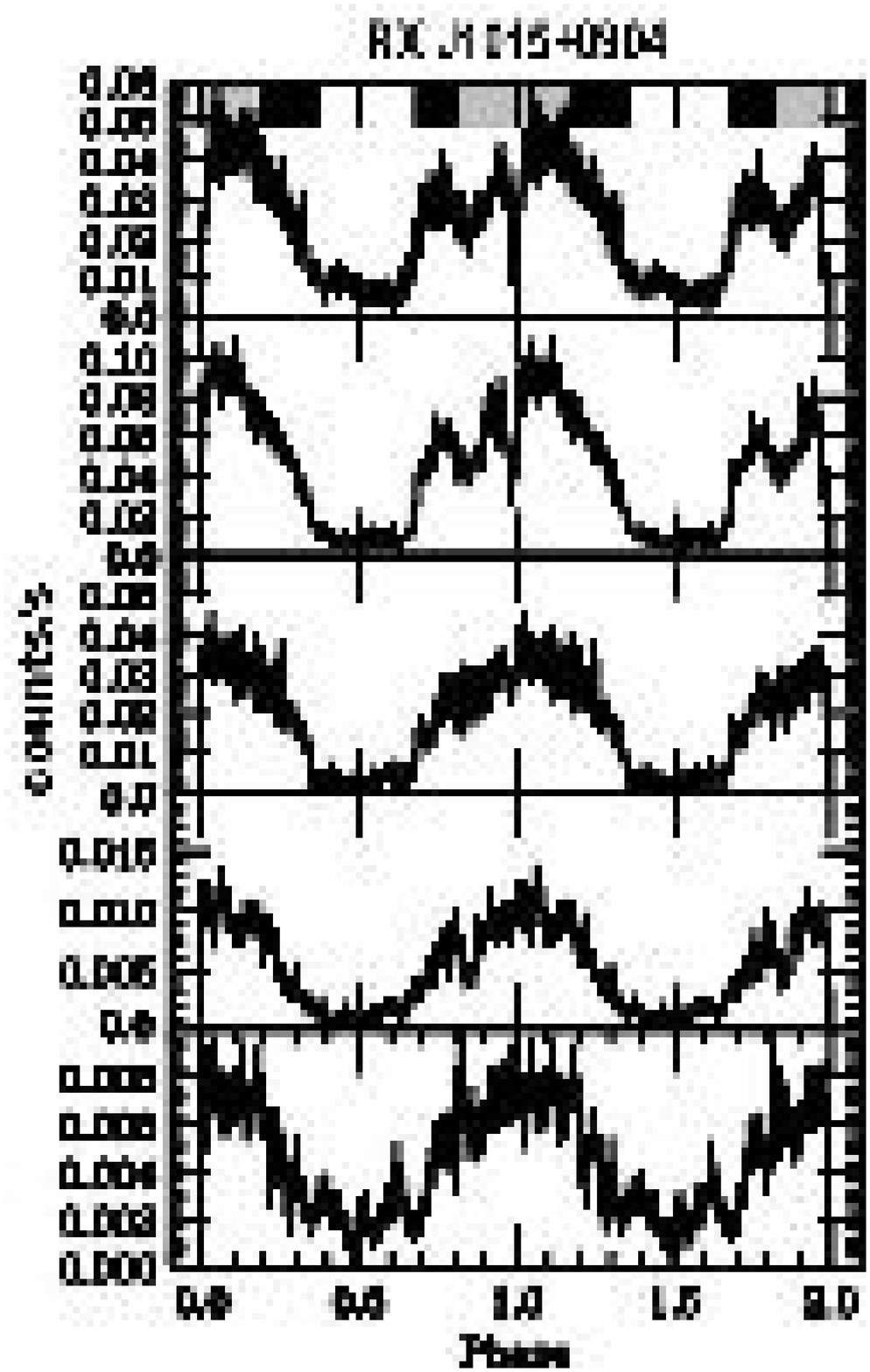}
\FigureFile(5.0cm,6.0cm){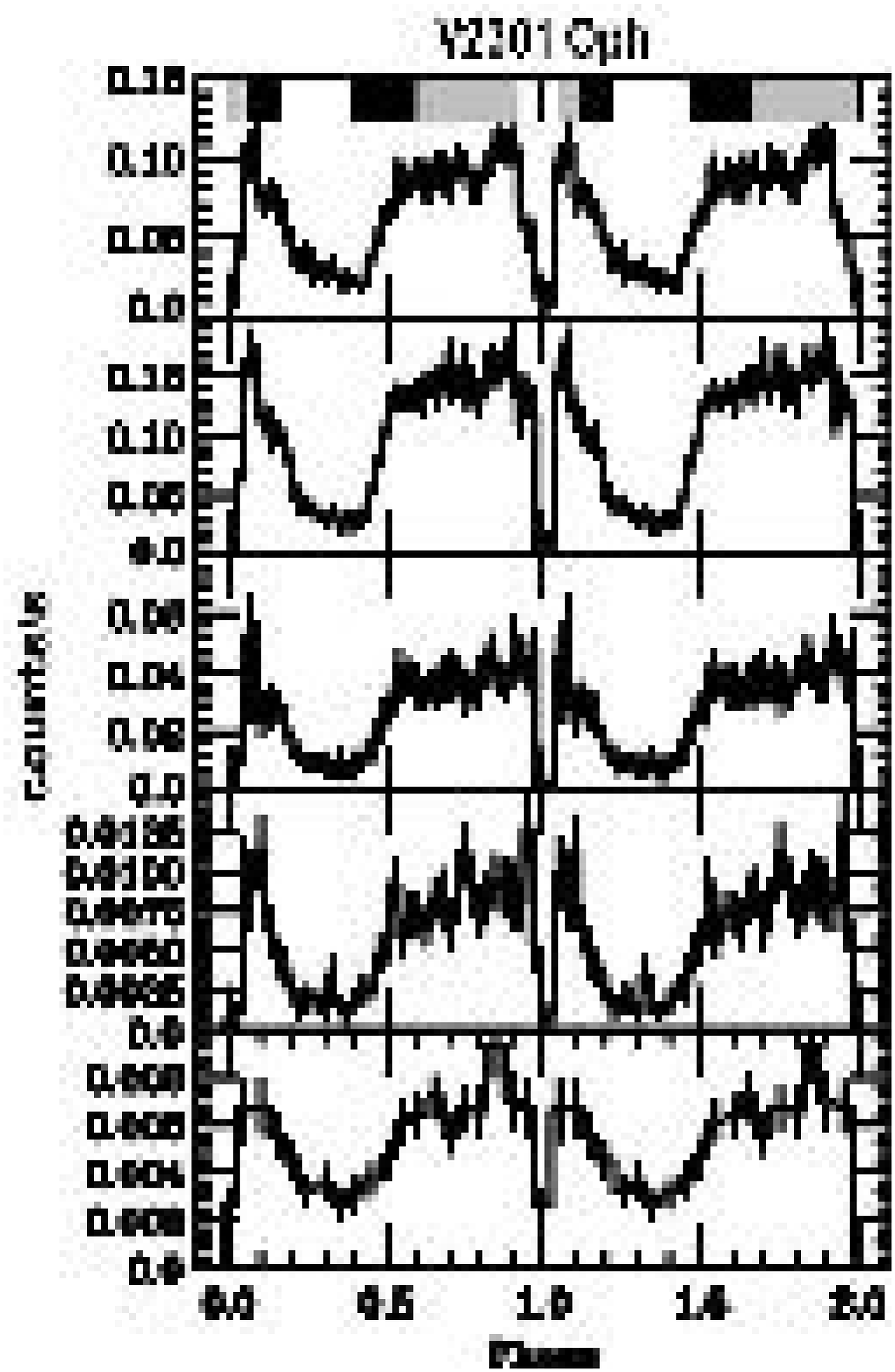}
\end{center}
\centerline{{\bf Fig.\ \ref{fig:other_polars_lc}} Continued.}
\end{figure*}

\begin{figure*}[hbt]
\begin{center}
\FigureFile(5.0cm,6.0cm){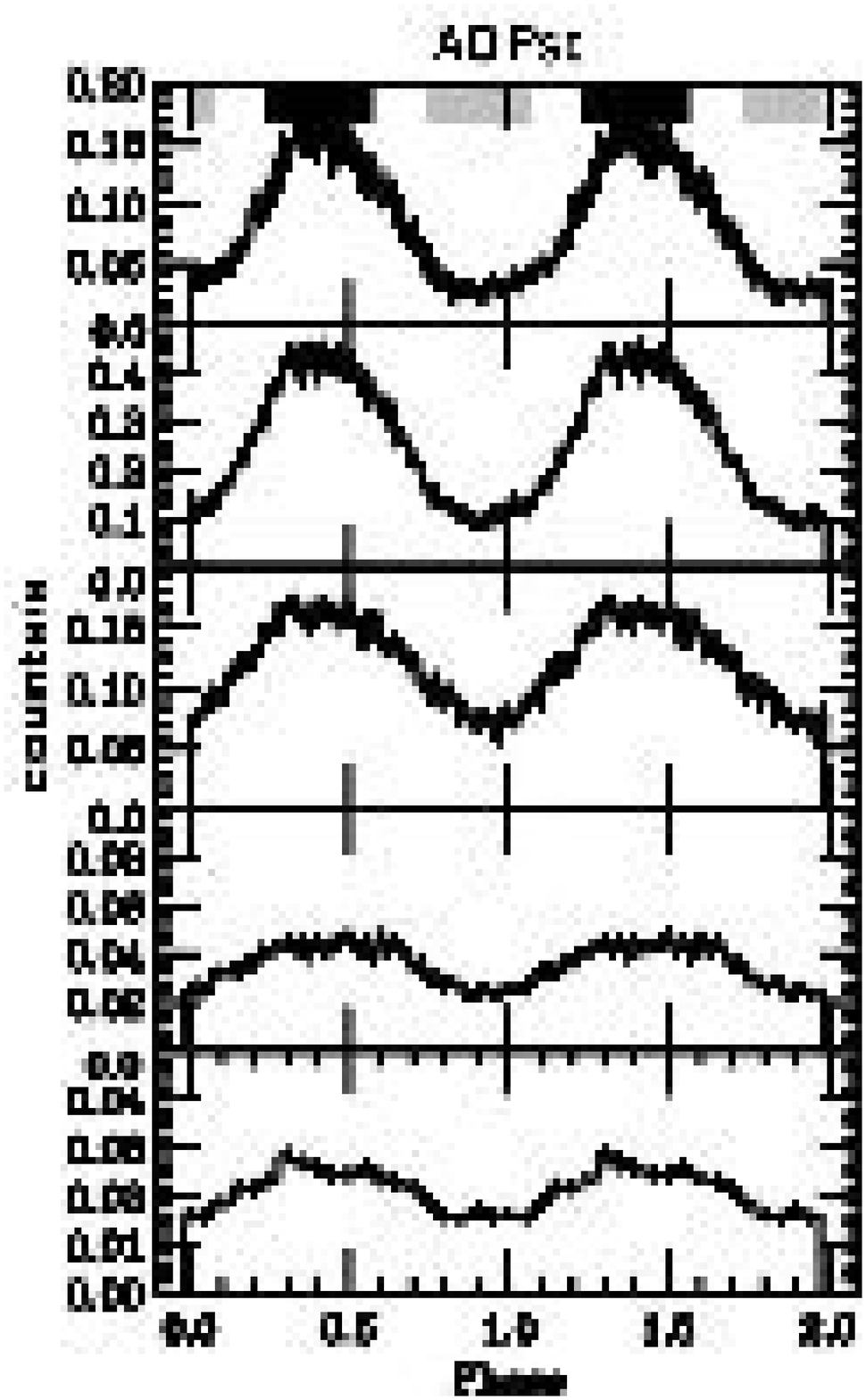}
\FigureFile(9.5cm,6cm){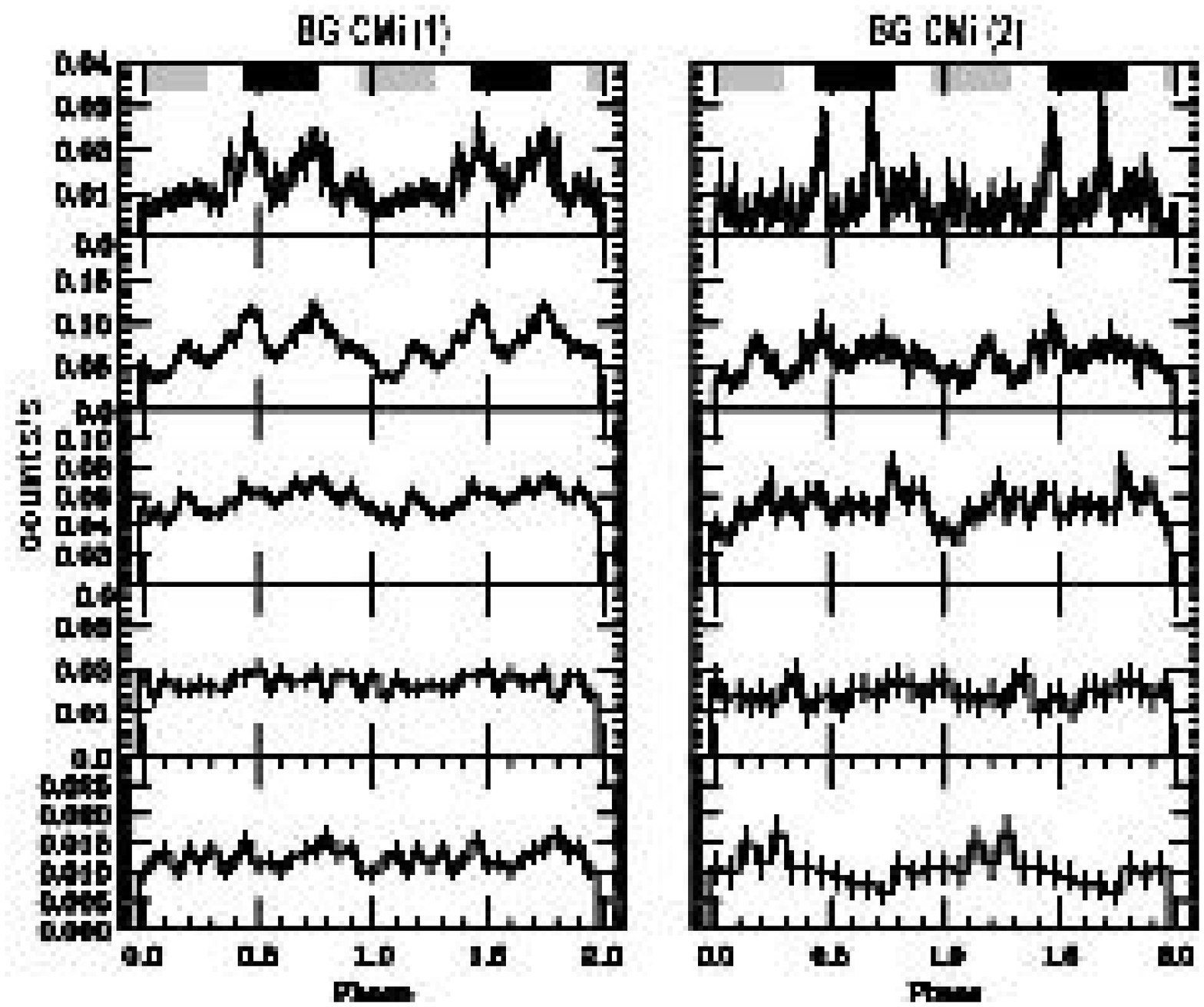}\\
\FigureFile(5.0cm,6.0cm){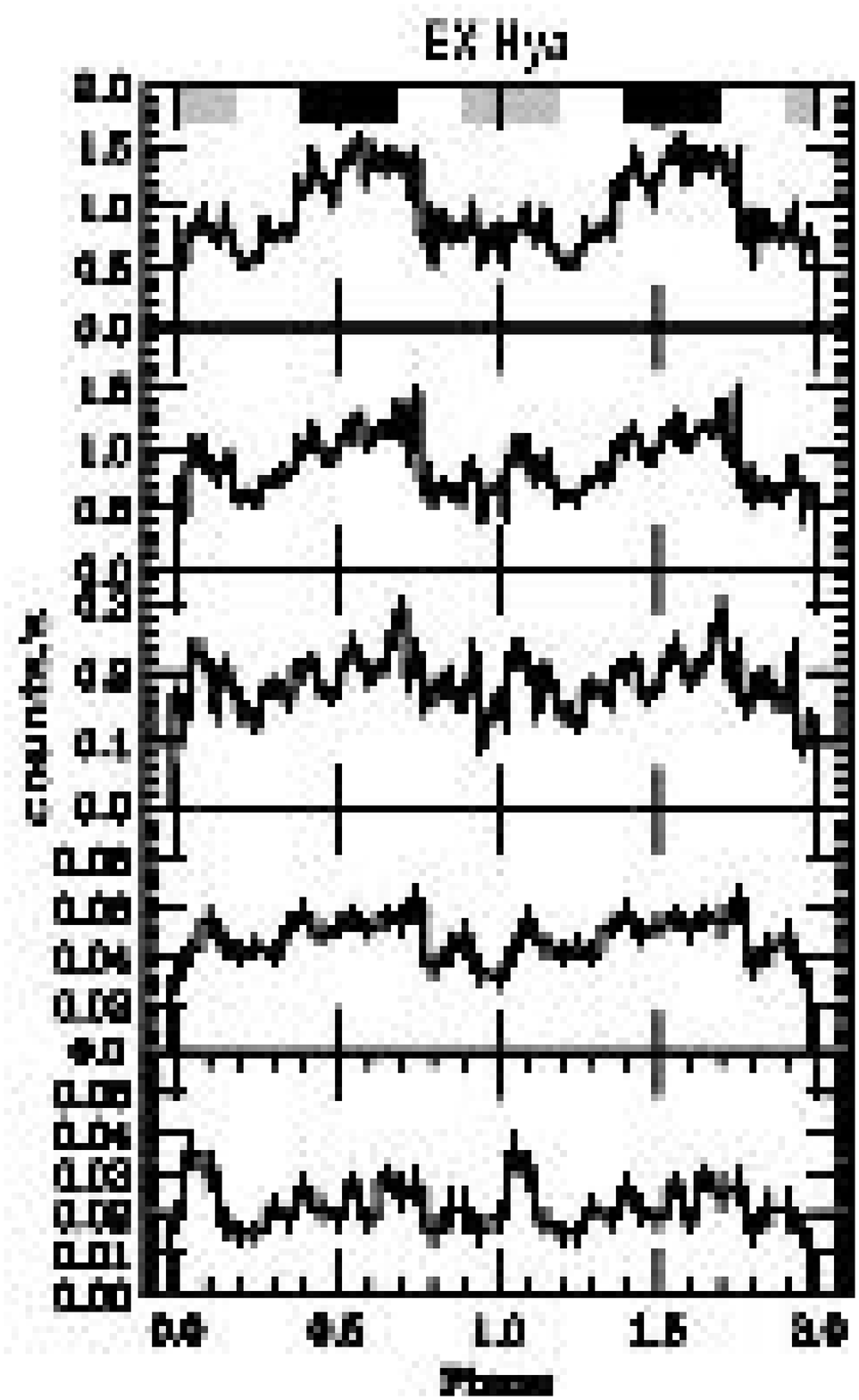}
\FigureFile(5.0cm,6.0cm){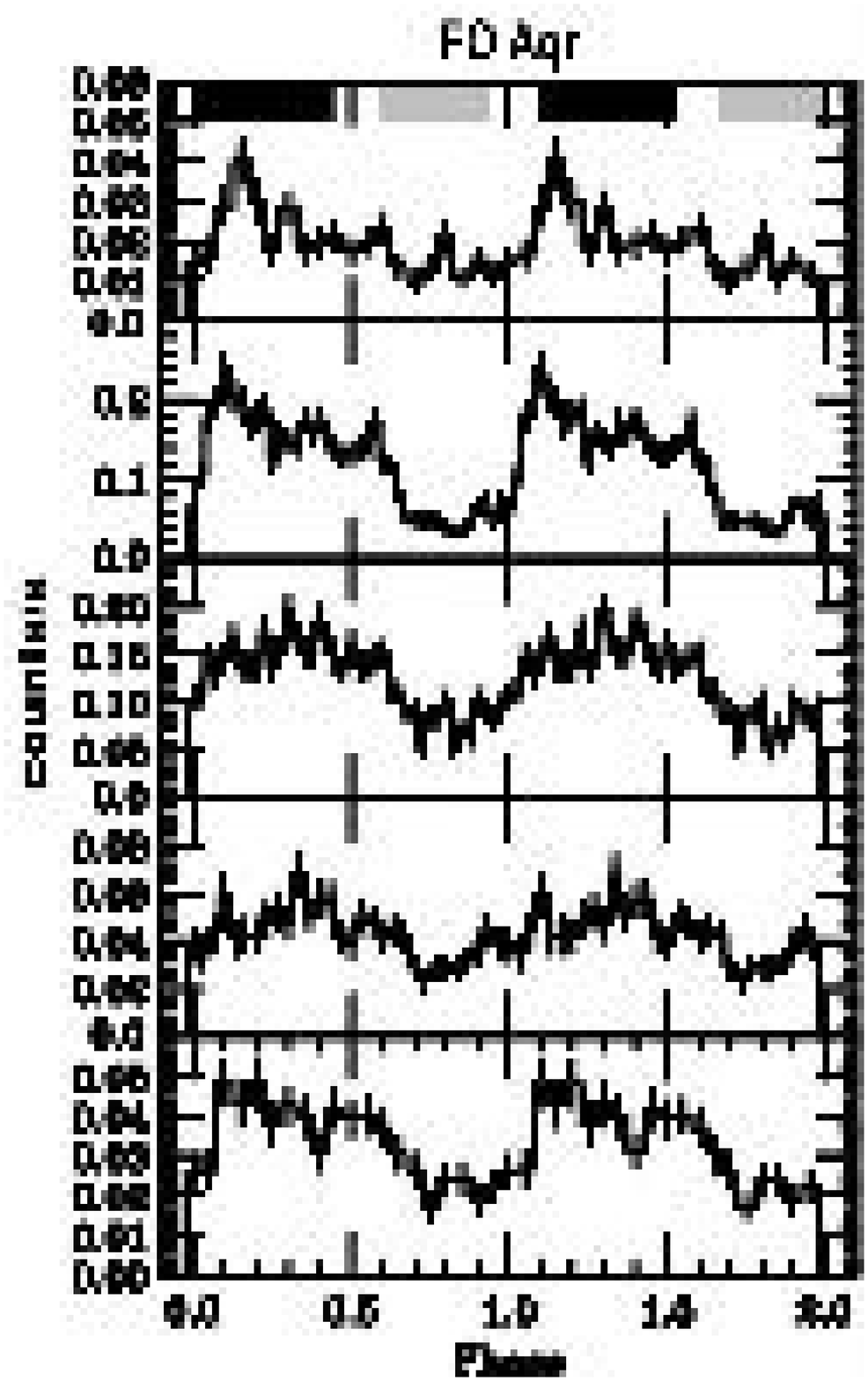}
\FigureFile(5.0cm,6.0cm){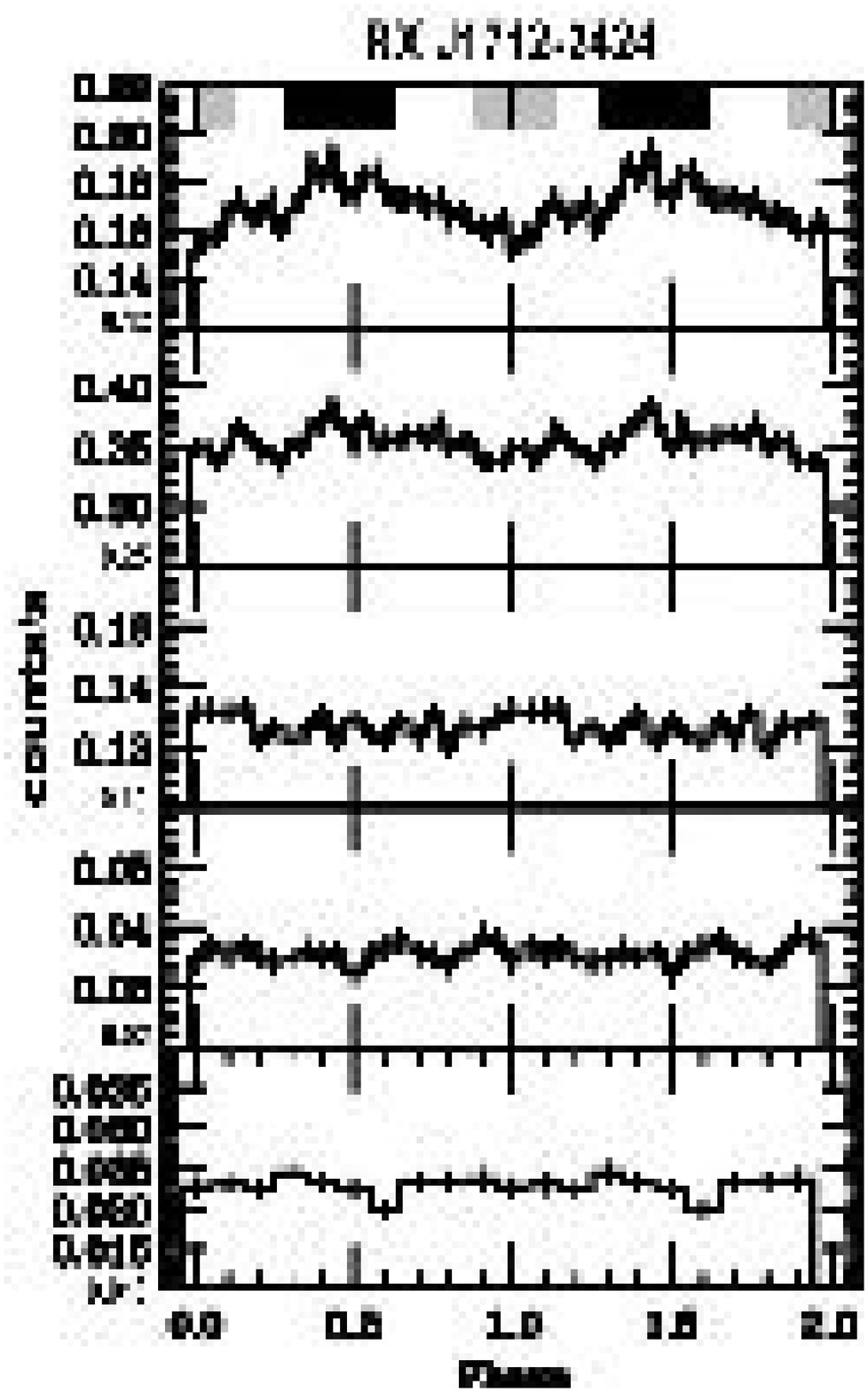}
\end{center}
\caption{Similar to figure \ref{fig:other_polars_lc}, but for the IPs.}
\label{fig:other_intermediate_polars_lc1}
\end{figure*}

\begin{figure*}[hbt]
\begin{center}
\FigureFile(9.5cm,6.0cm){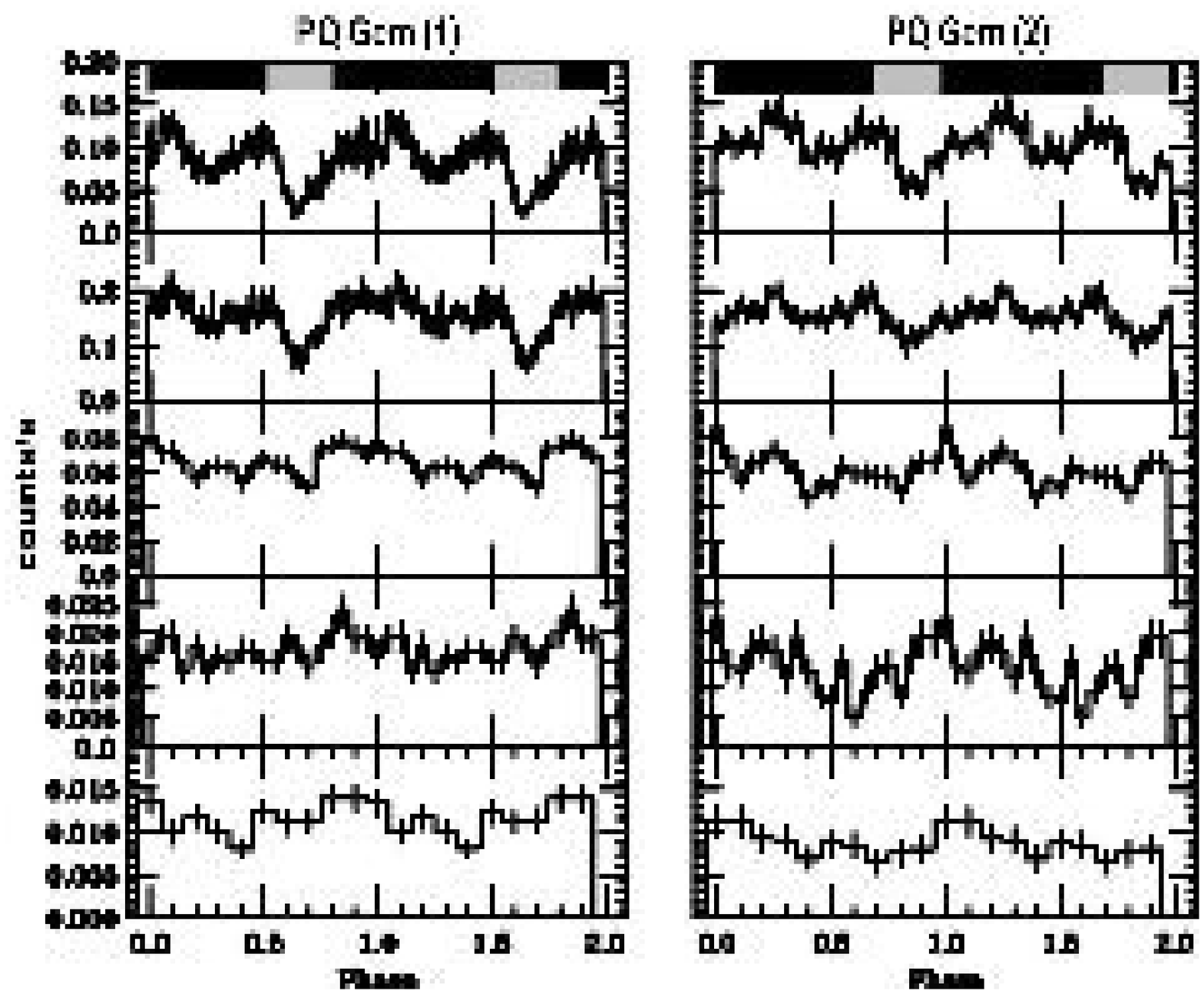}
\FigureFile(5.0cm,6.0cm){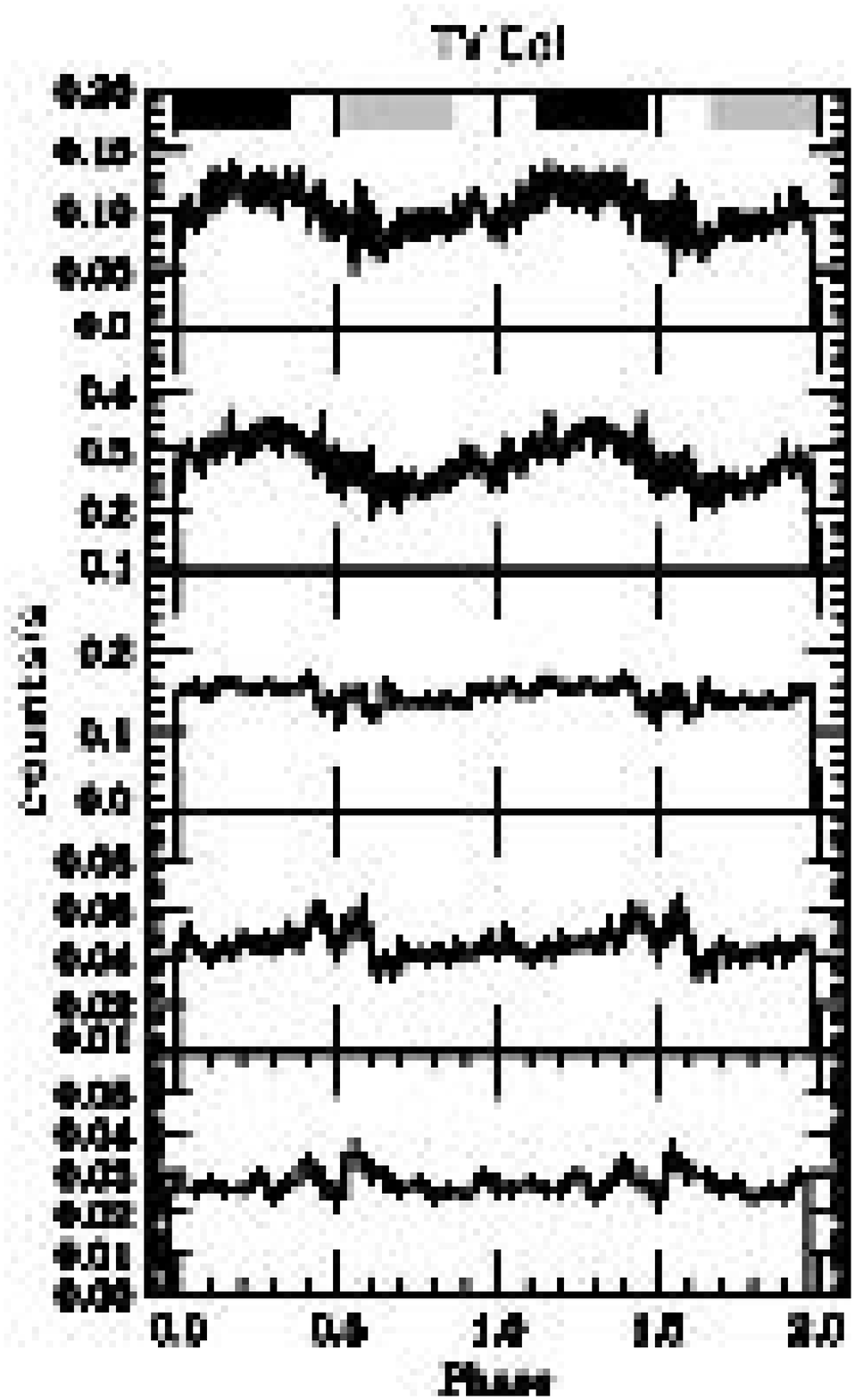}\\
\FigureFile(5.0cm,6.0cm){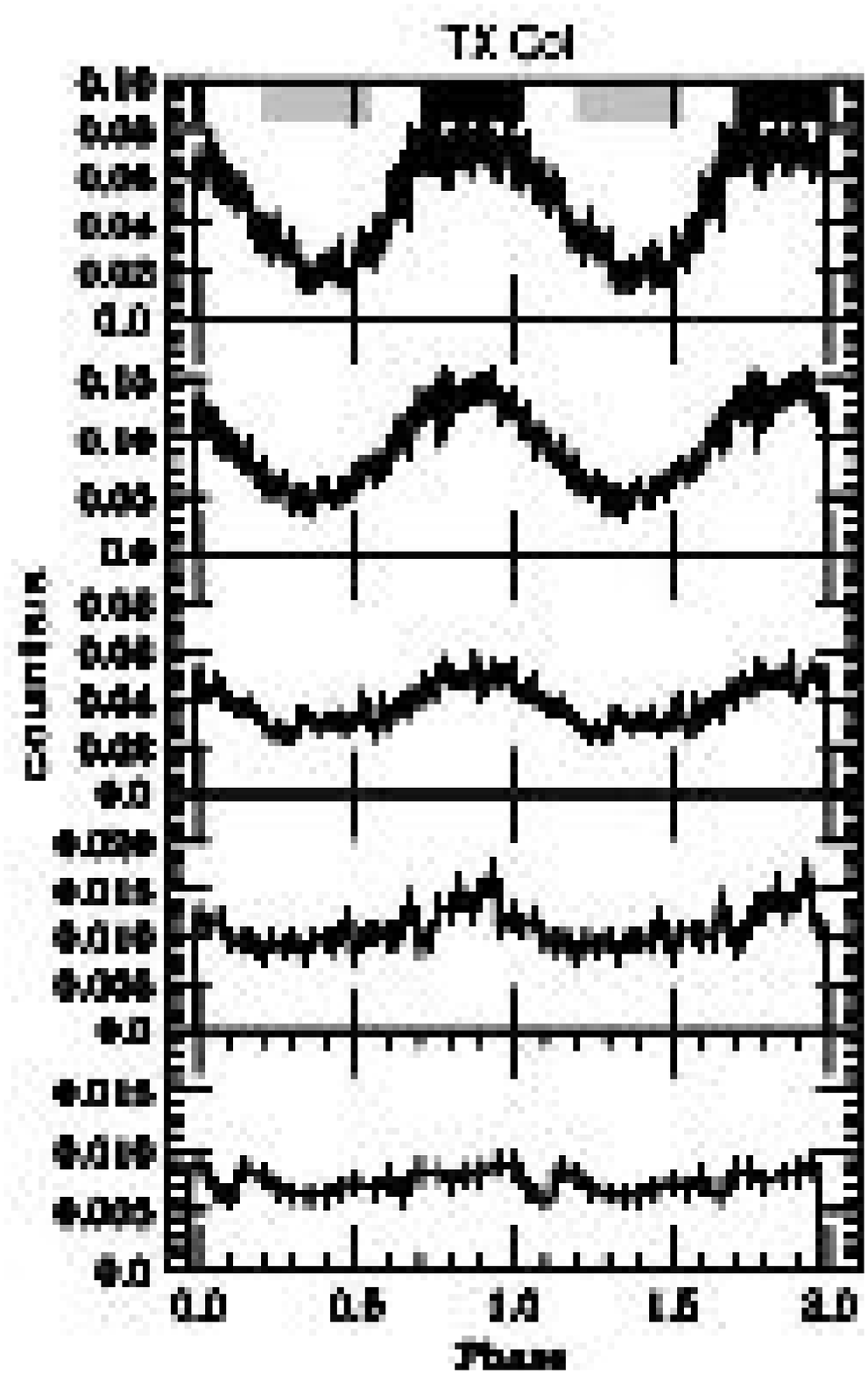}
\FigureFile(5.0cm,6.0cm){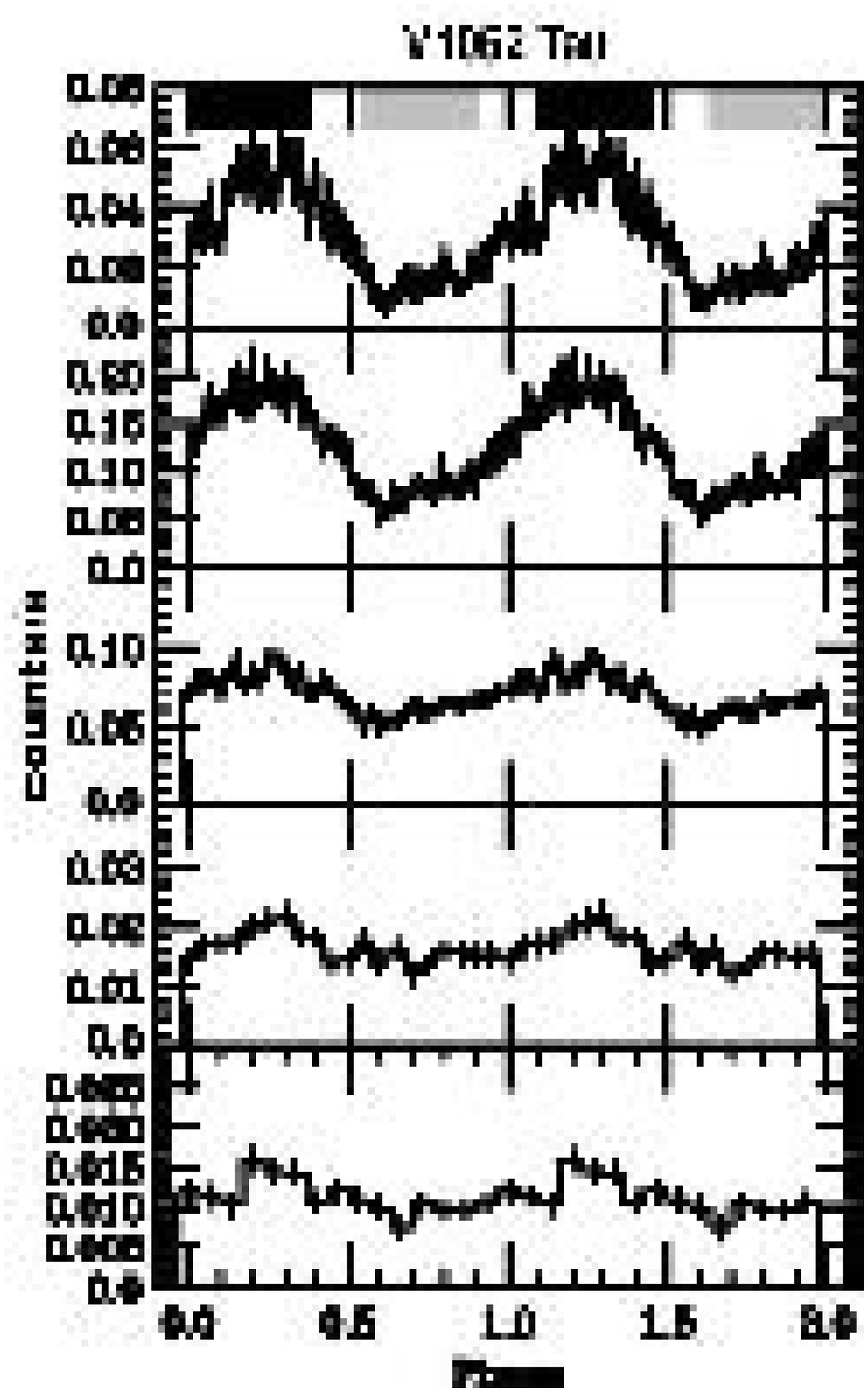}
\FigureFile(5.0cm,6.0cm){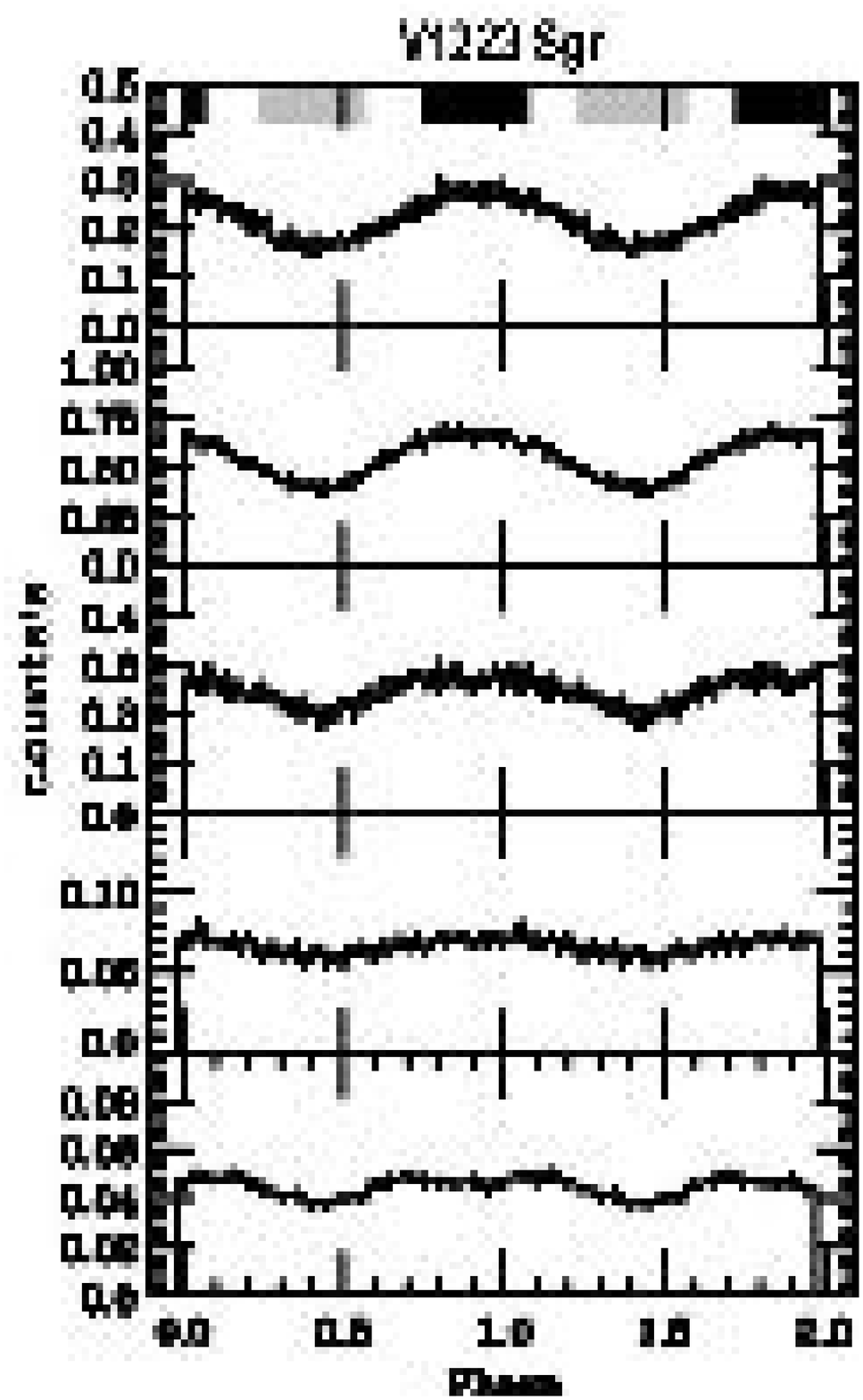} 
\end{center}
\centerline{{\bf Fig.\ \ref{fig:other_intermediate_polars_lc1}} Continued.}
\end{figure*}

\begin{figure*}[hbt]
\begin{center}
\FigureFile(9.5cm,6.0cm) {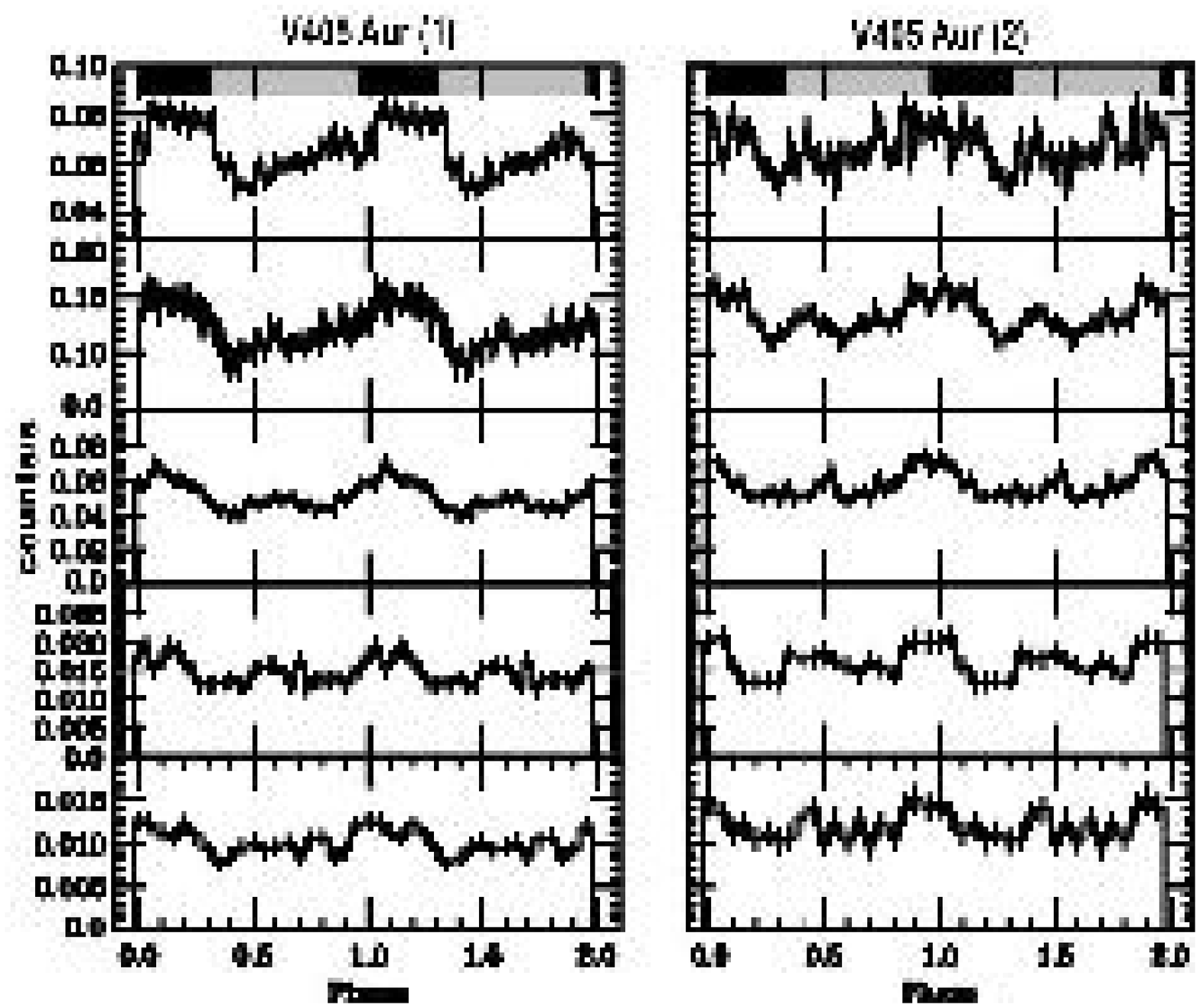}
\FigureFile(15.0cm,6.0cm){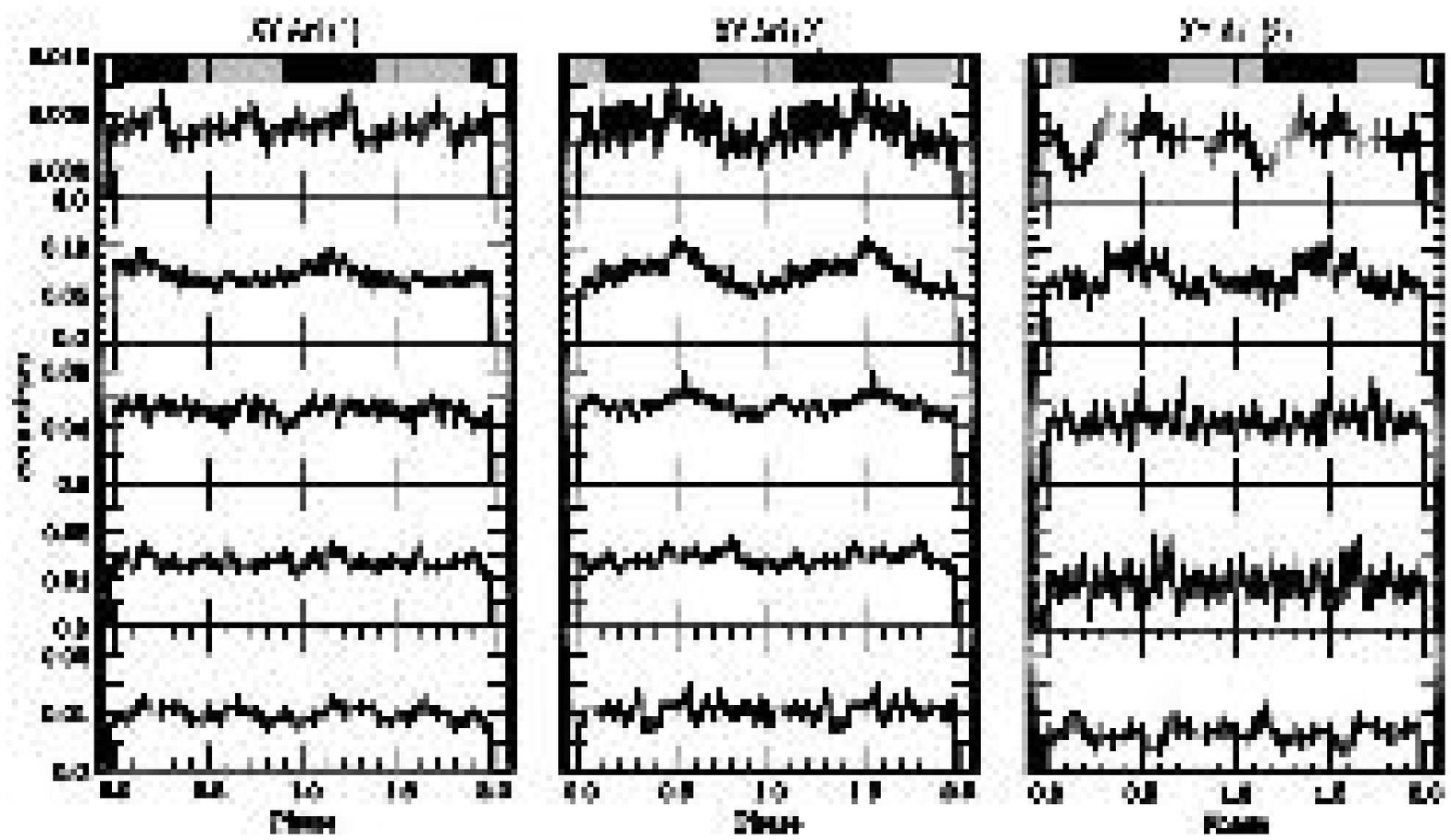}
\end{center}
\centerline{{\bf Fig.\ \ref{fig:other_intermediate_polars_lc1}} Continued.}
\end{figure*}

\begin{figure*}[hbt]
\begin{center}
\FigureFile(7.5cm,12cm){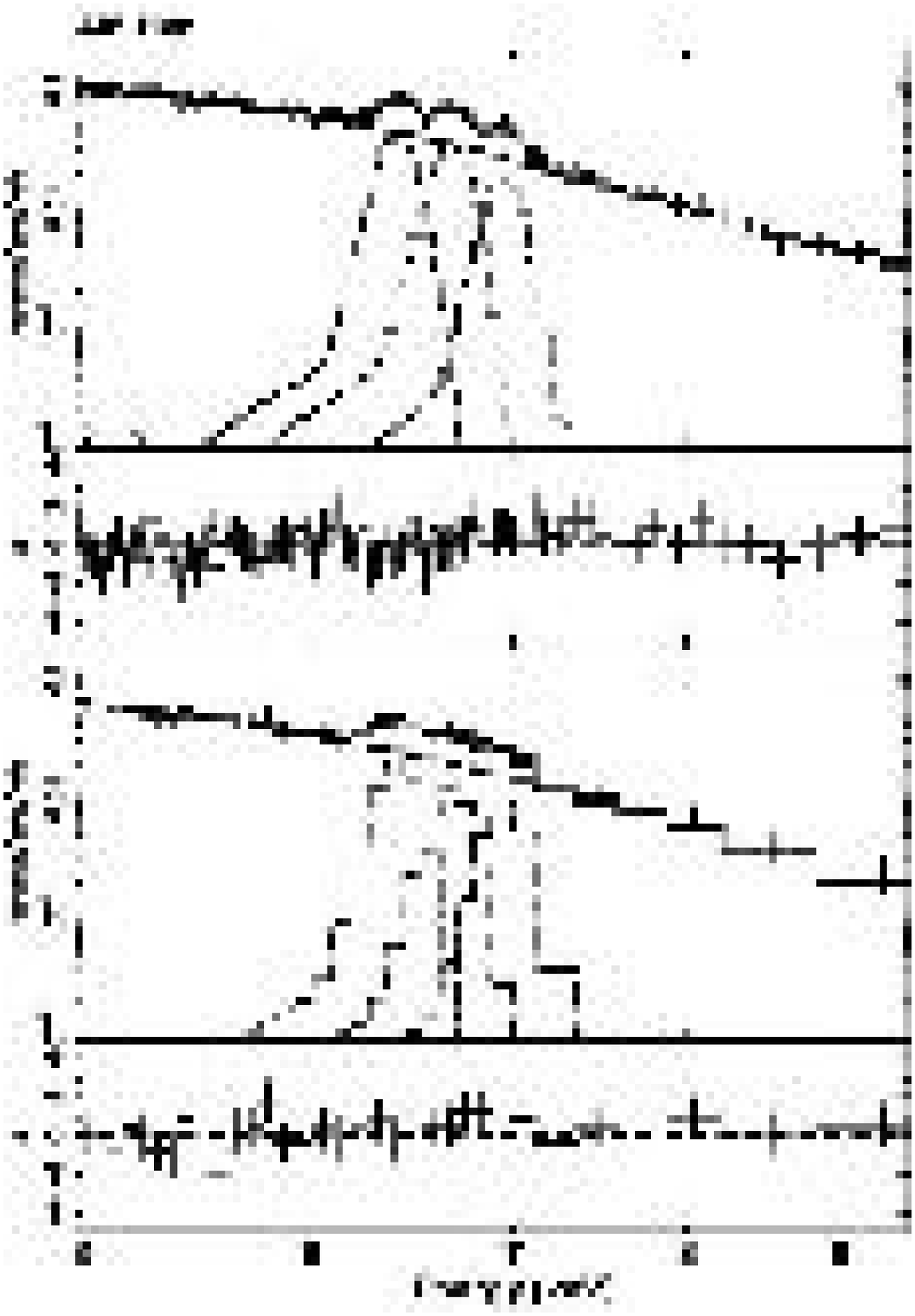}
\FigureFile(7.5cm,12cm){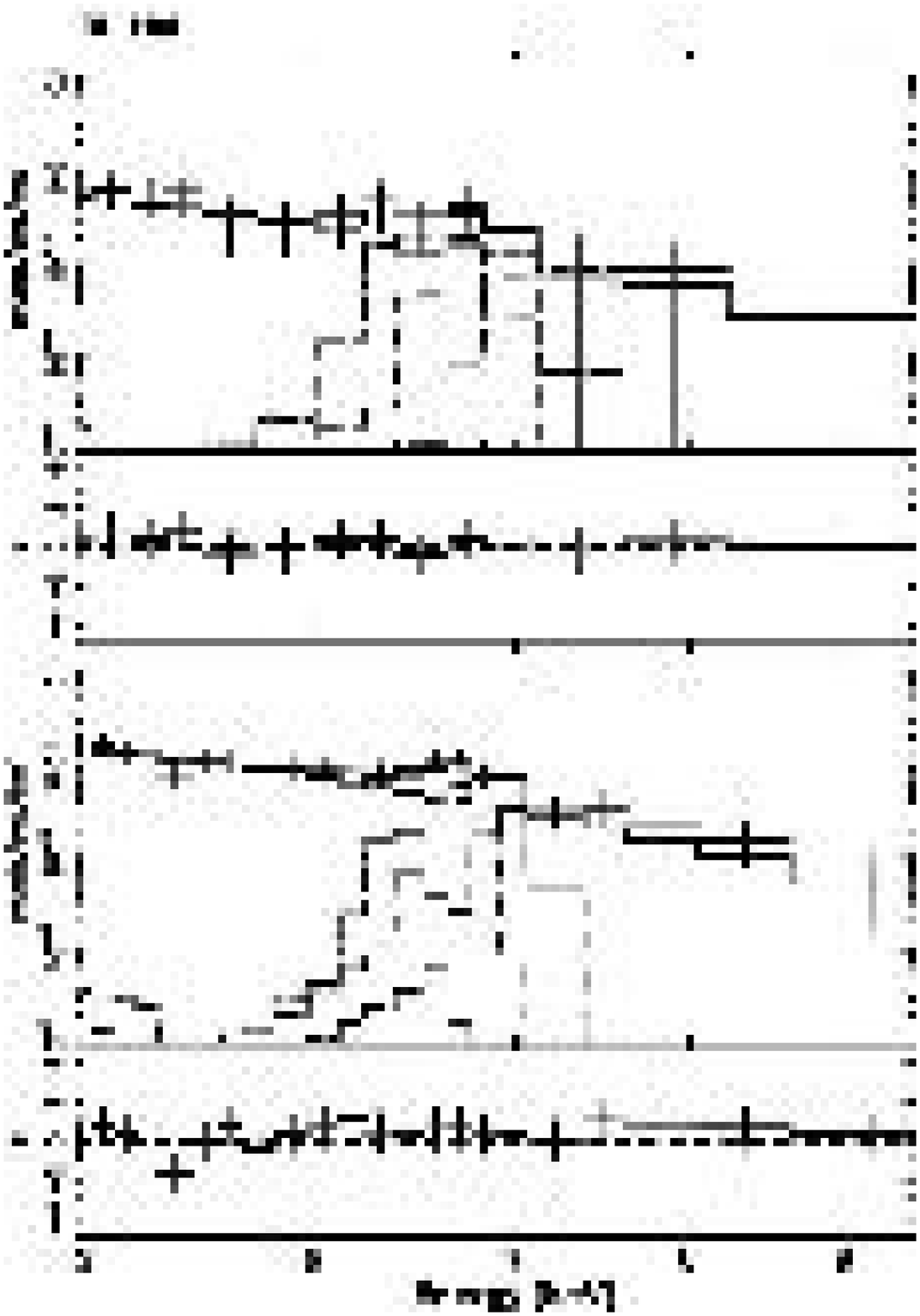}
\FigureFile(7.5cm,12cm){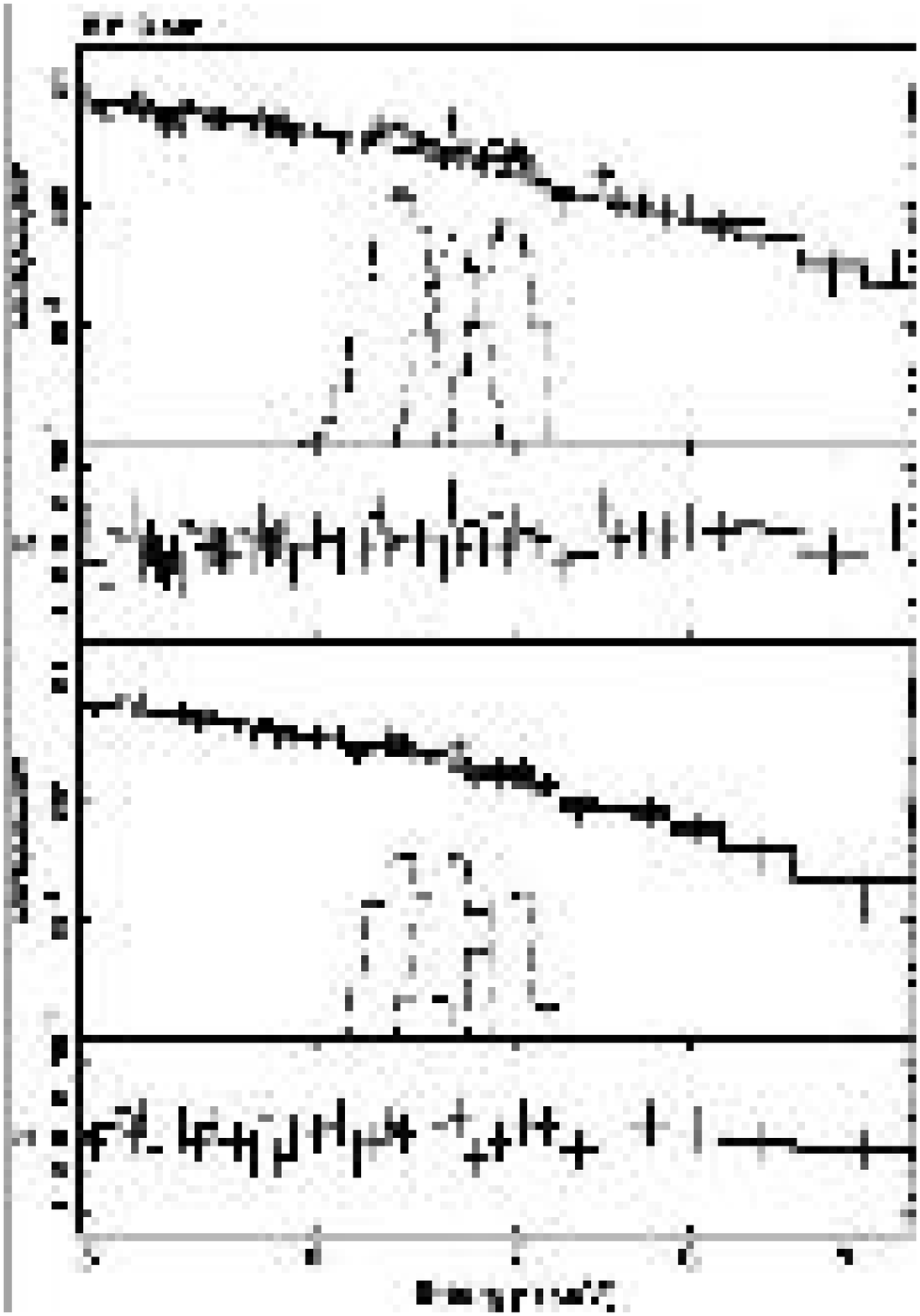}
\FigureFile(7.5cm,12cm){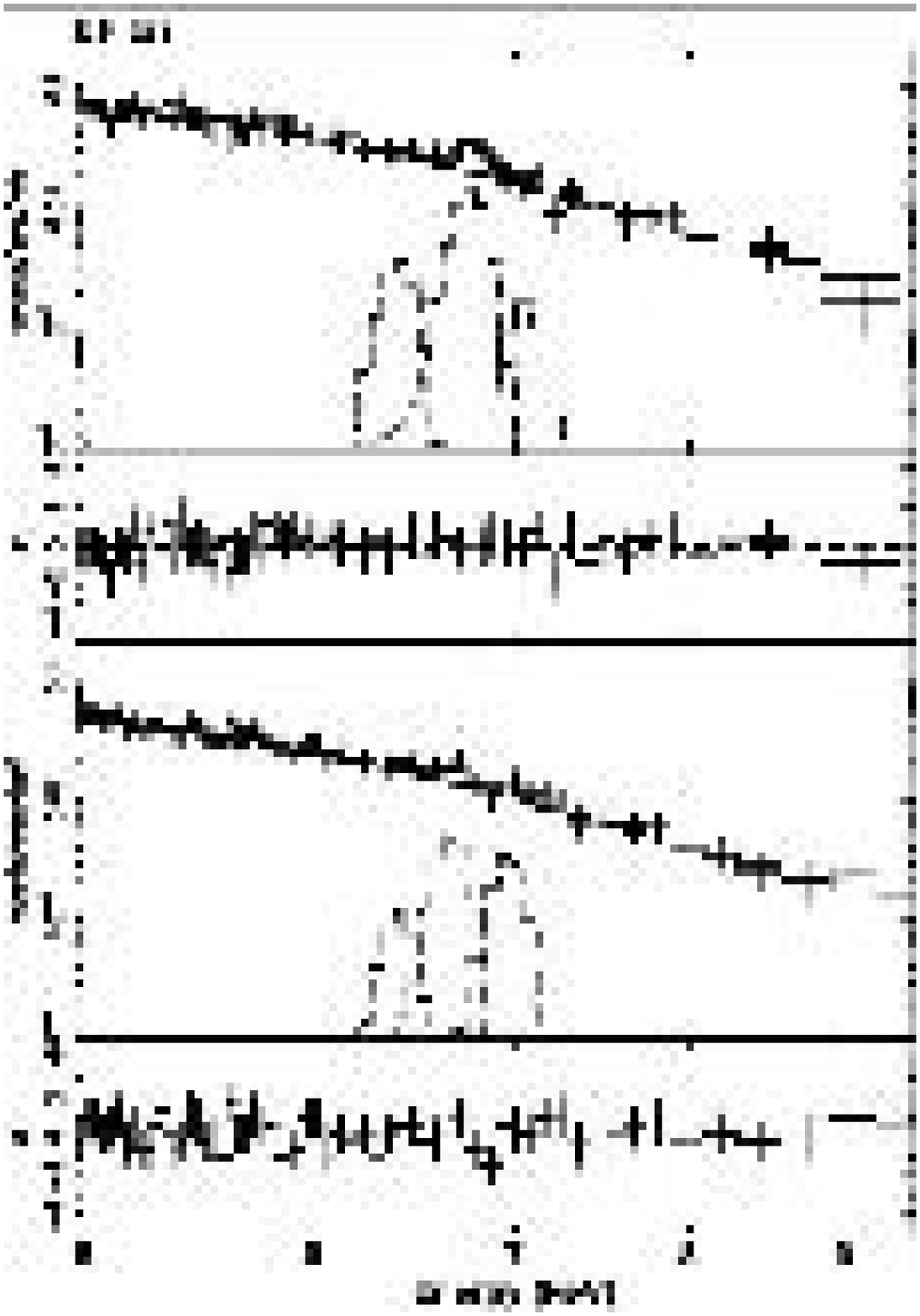}
\end{center}
\caption{Phase resolved spectra near the Fe energy band of 
the polars in our sample, acquired with {\it ASCA}. 
Top panels show the pole-on spectra, and bottom panels the side-on spectra.
Only the SIS data are shown in this figure.
The best fit models, obtained by simultaneous fitting
the GIS and SIS data, are also plotted, 
after convolving with the detector response.}
\label{fig:polars_spec}
\end{figure*}

\begin{figure*}[hbt]
\begin{center}
\FigureFile(7.5cm,12cm){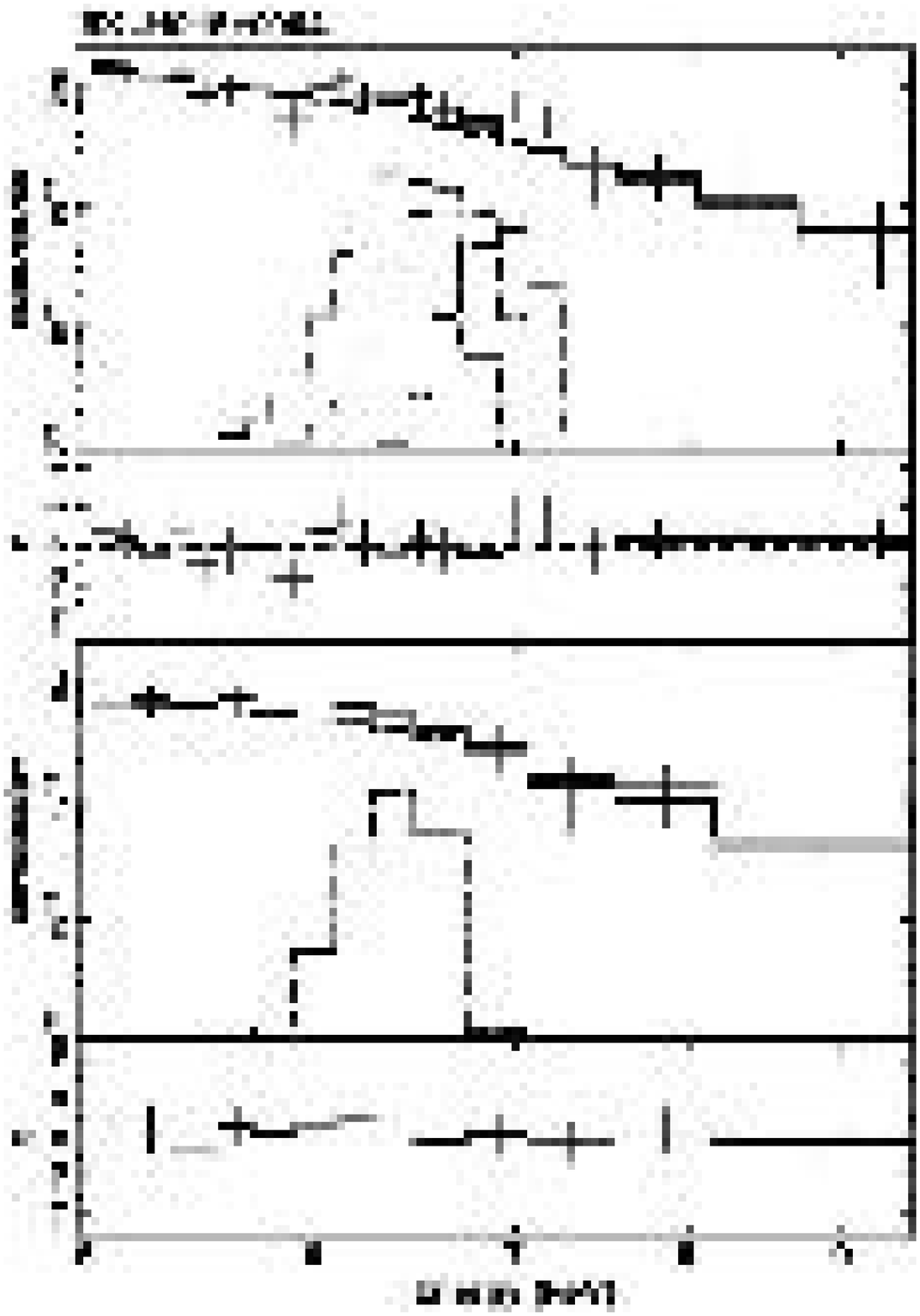}
\FigureFile(7.5cm,12cm){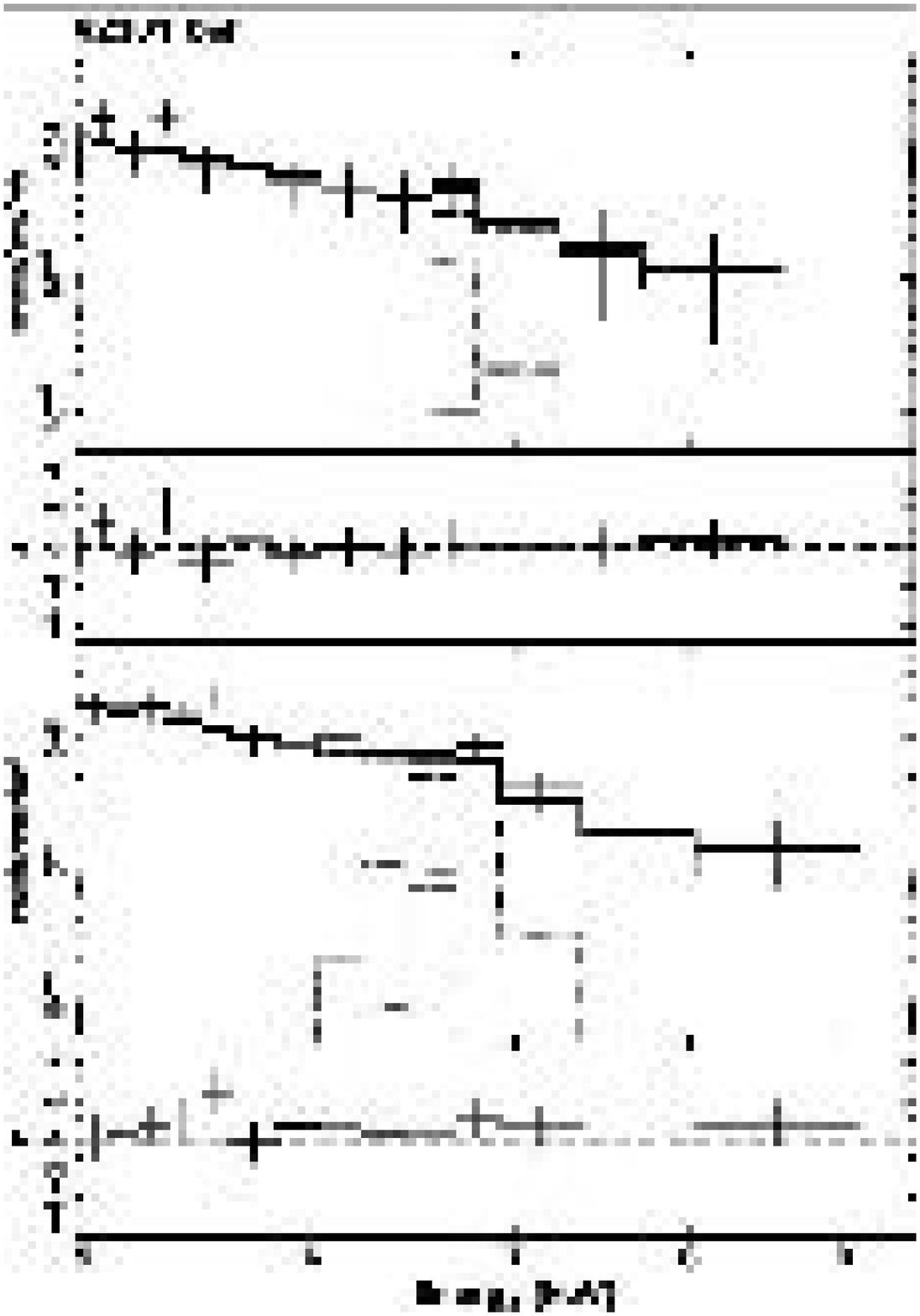}
\end{center}
\centerline{{\bf Fig.\ \ref{fig:polars_spec}} Continued.}
\end{figure*}

\begin{figure*}[hbt]
\begin{center}
\FigureFile(7.5cm,12cm){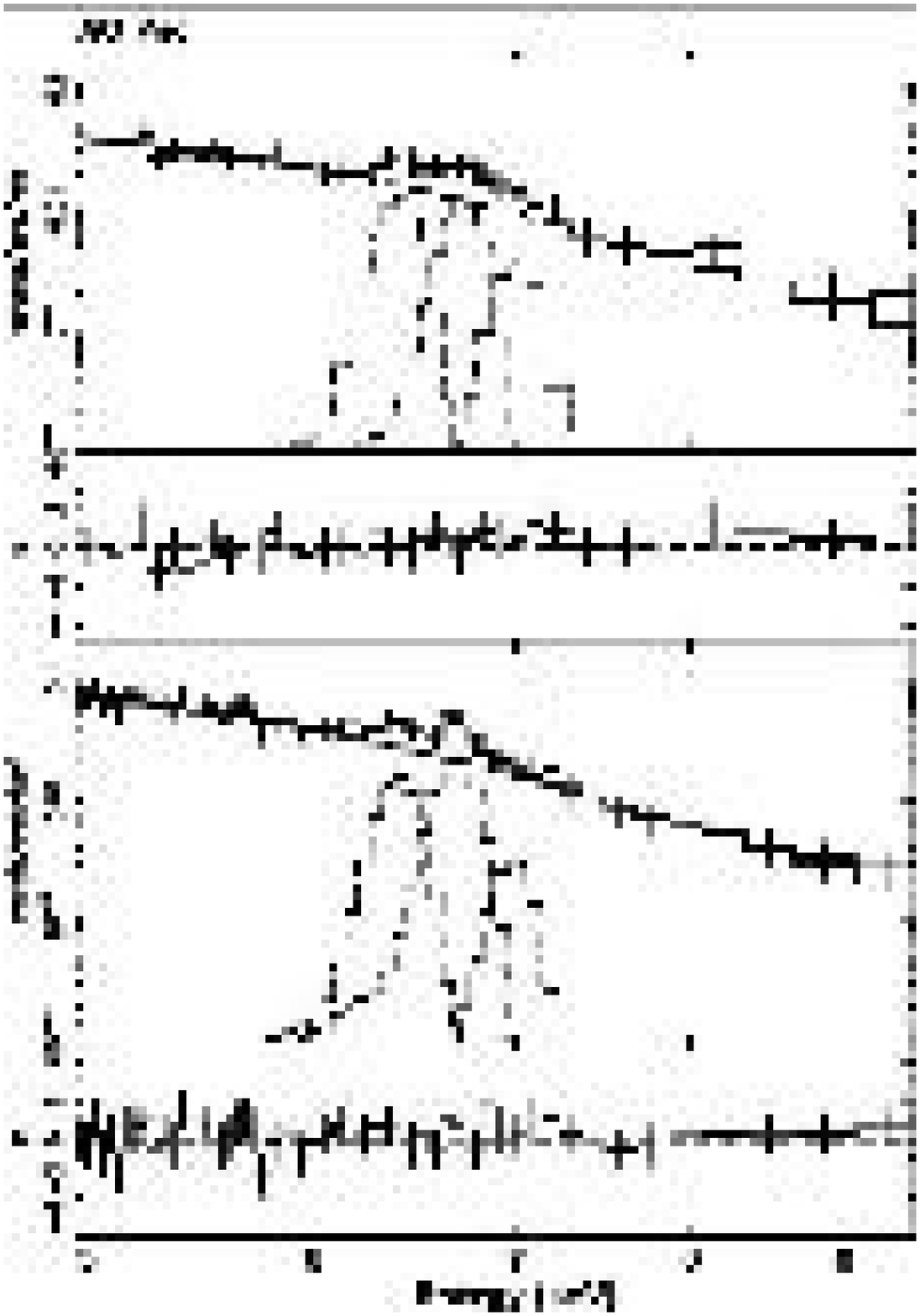}
\FigureFile(7.5cm,12cm){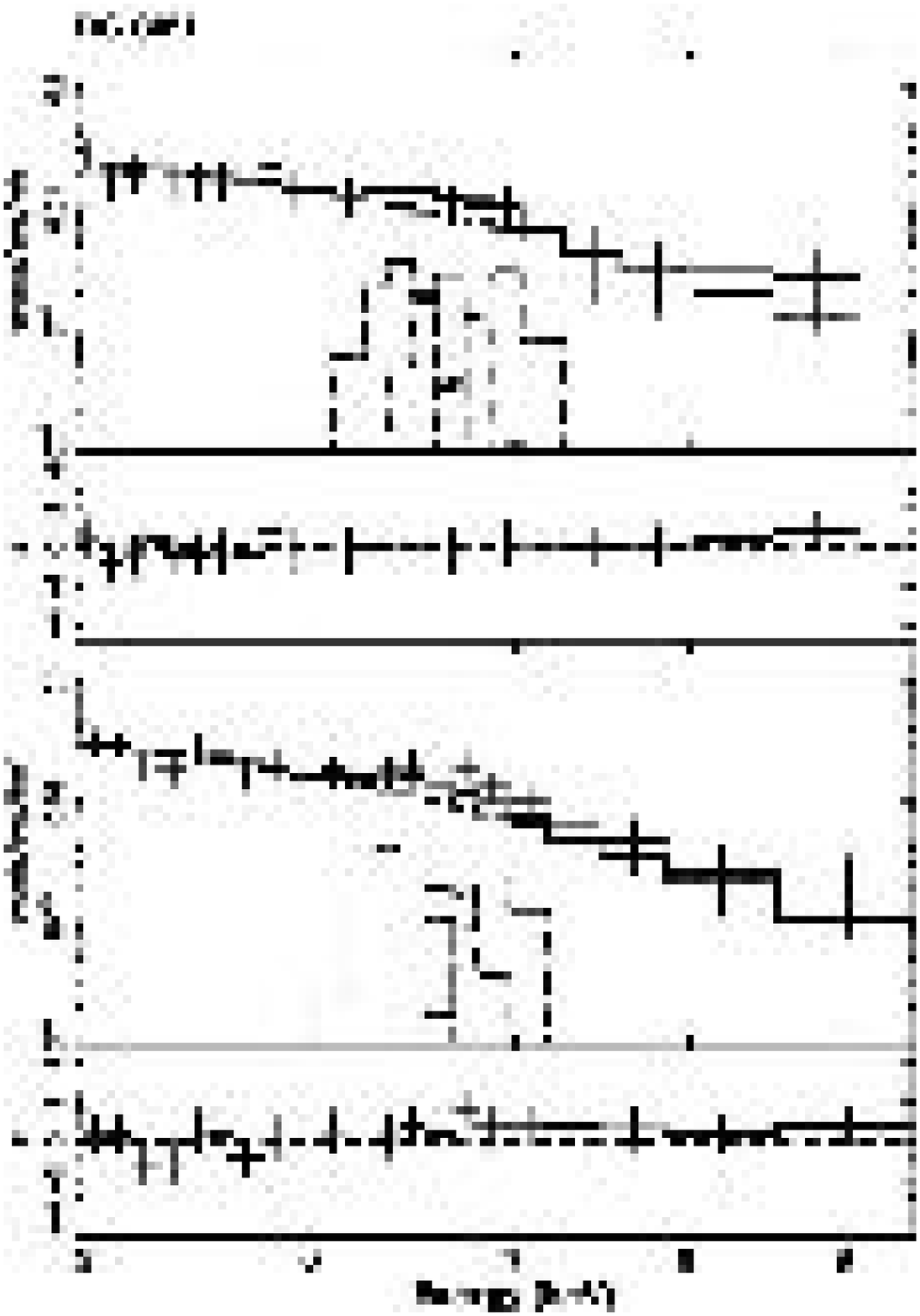}
\FigureFile(7.5cm,12cm){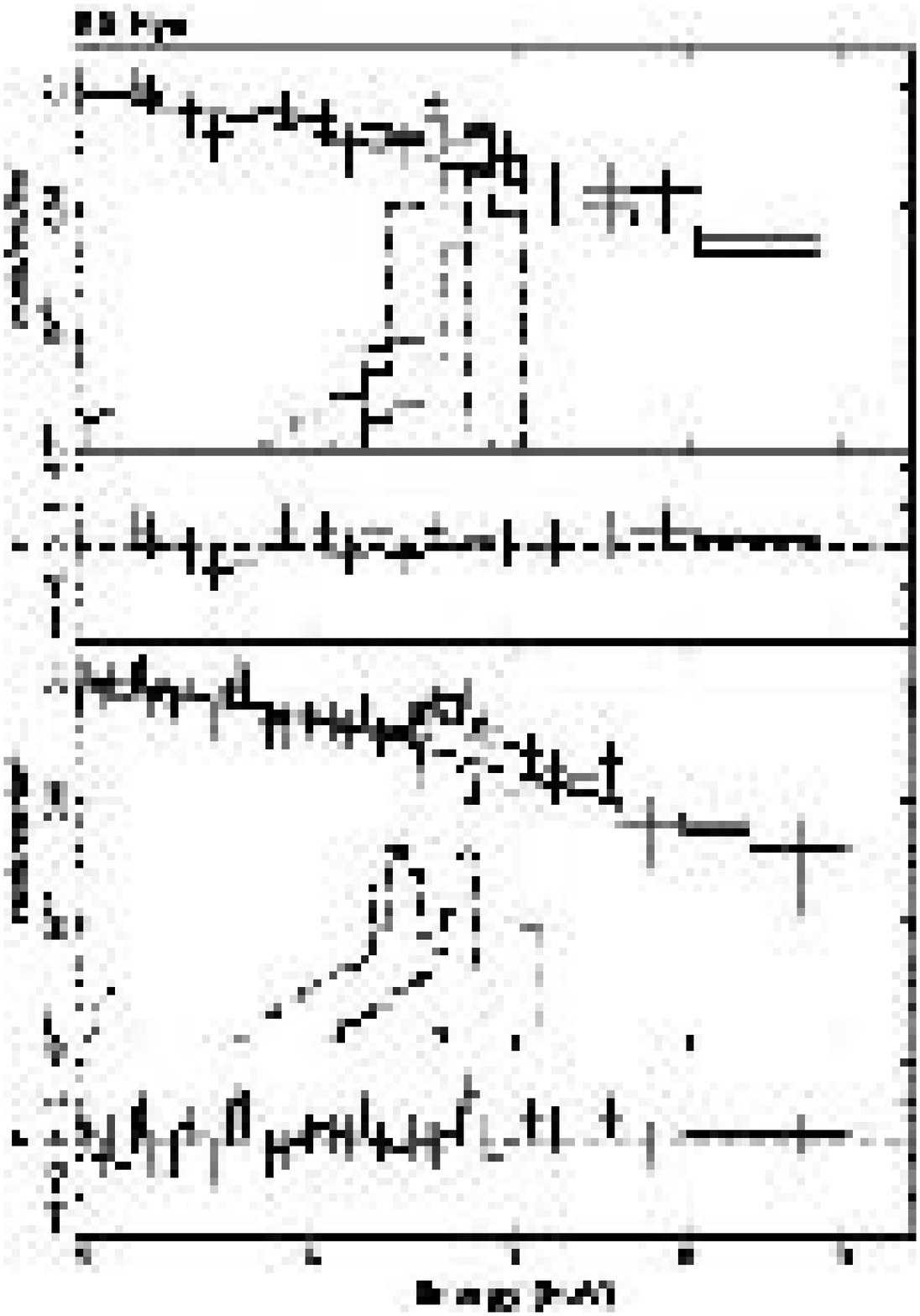}
\FigureFile(7.5cm,12cm){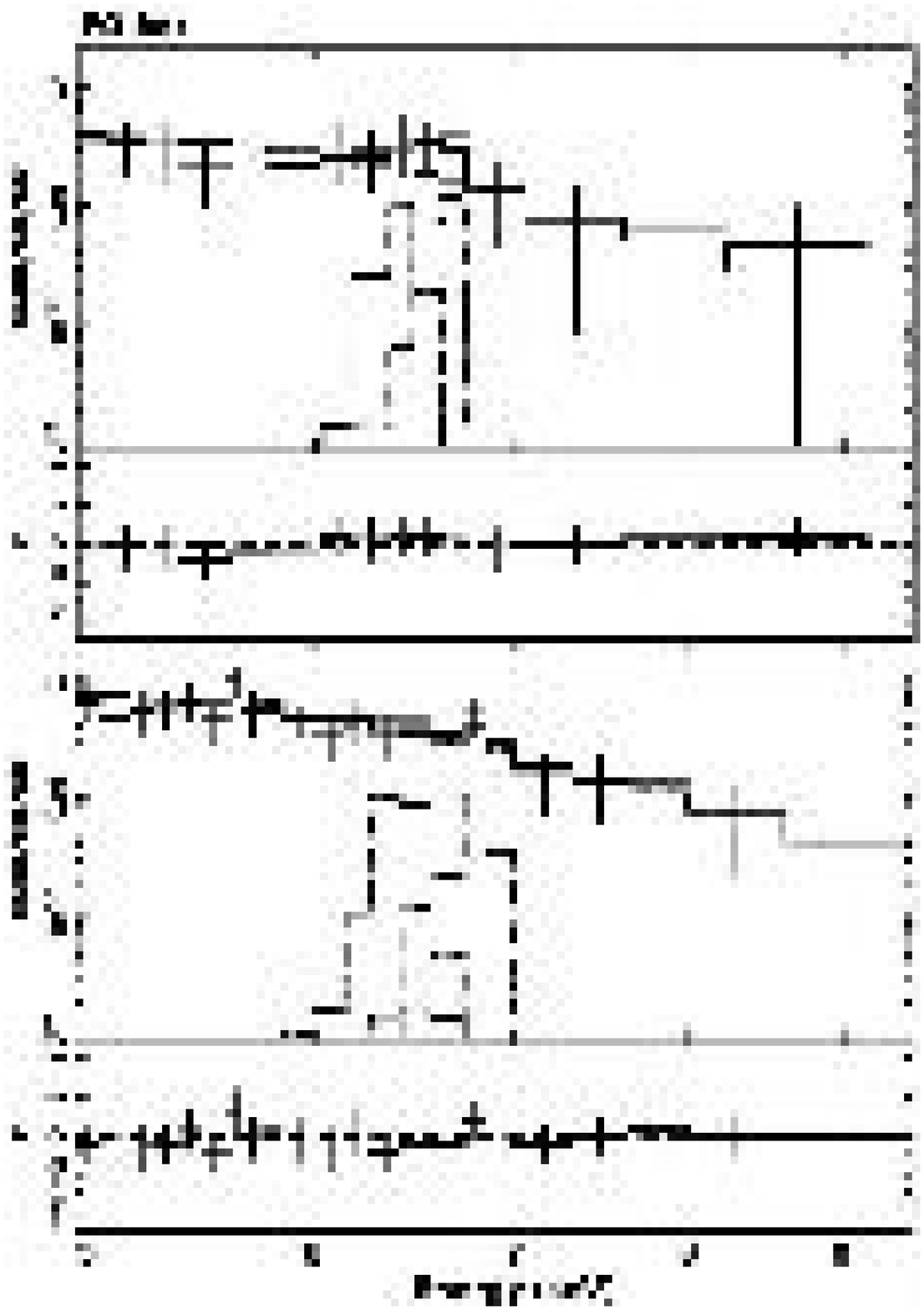}
\end{center}
\caption{Similar to figure \ref{fig:polars_spec},
but for the IPs in our sample. }
\label{fig:intermediate_polars_spec}
\end{figure*}

\begin{figure*}[hbt]
\begin{center}
\FigureFile(7.5cm,12cm){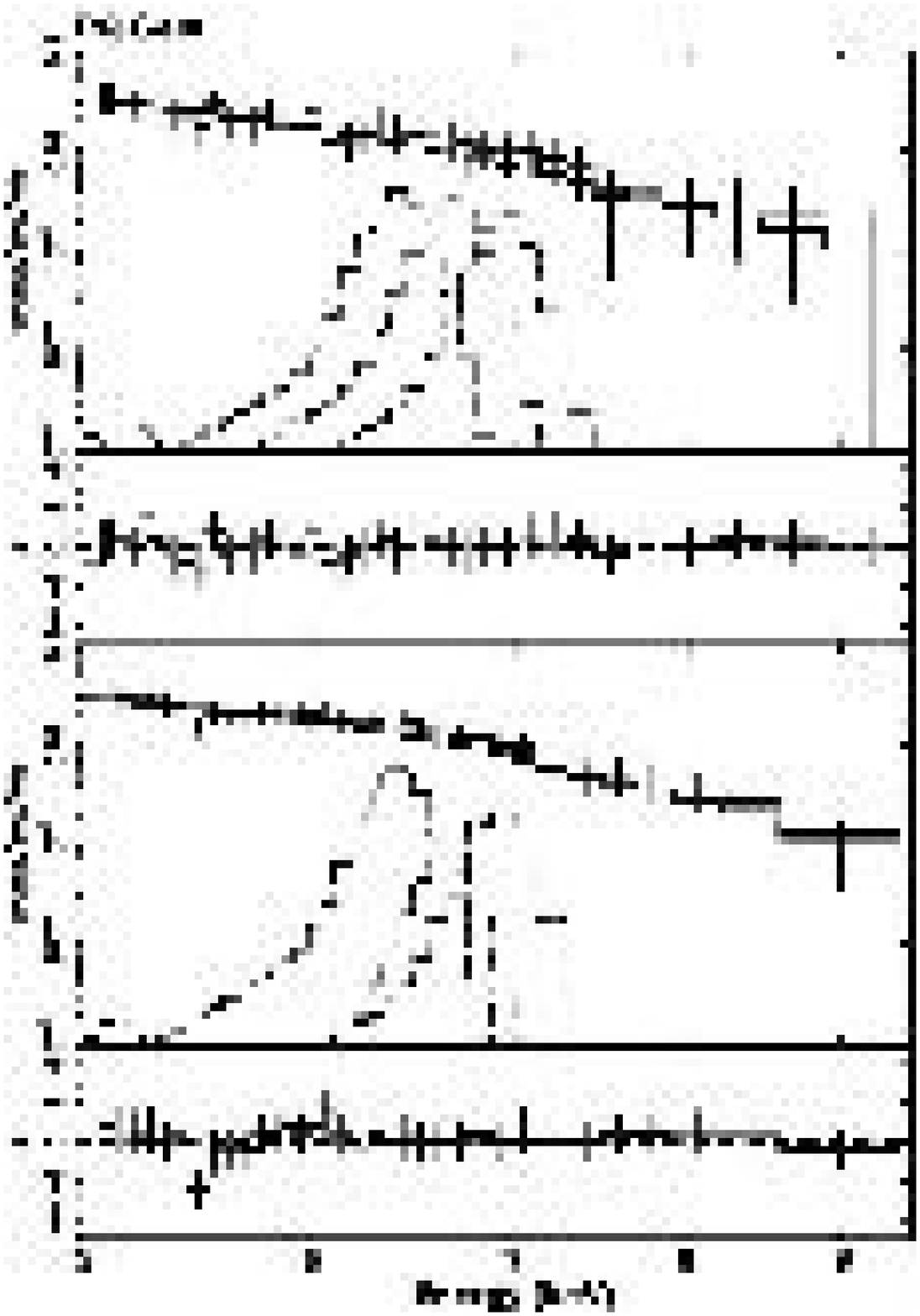}
\FigureFile(7.5cm,12cm){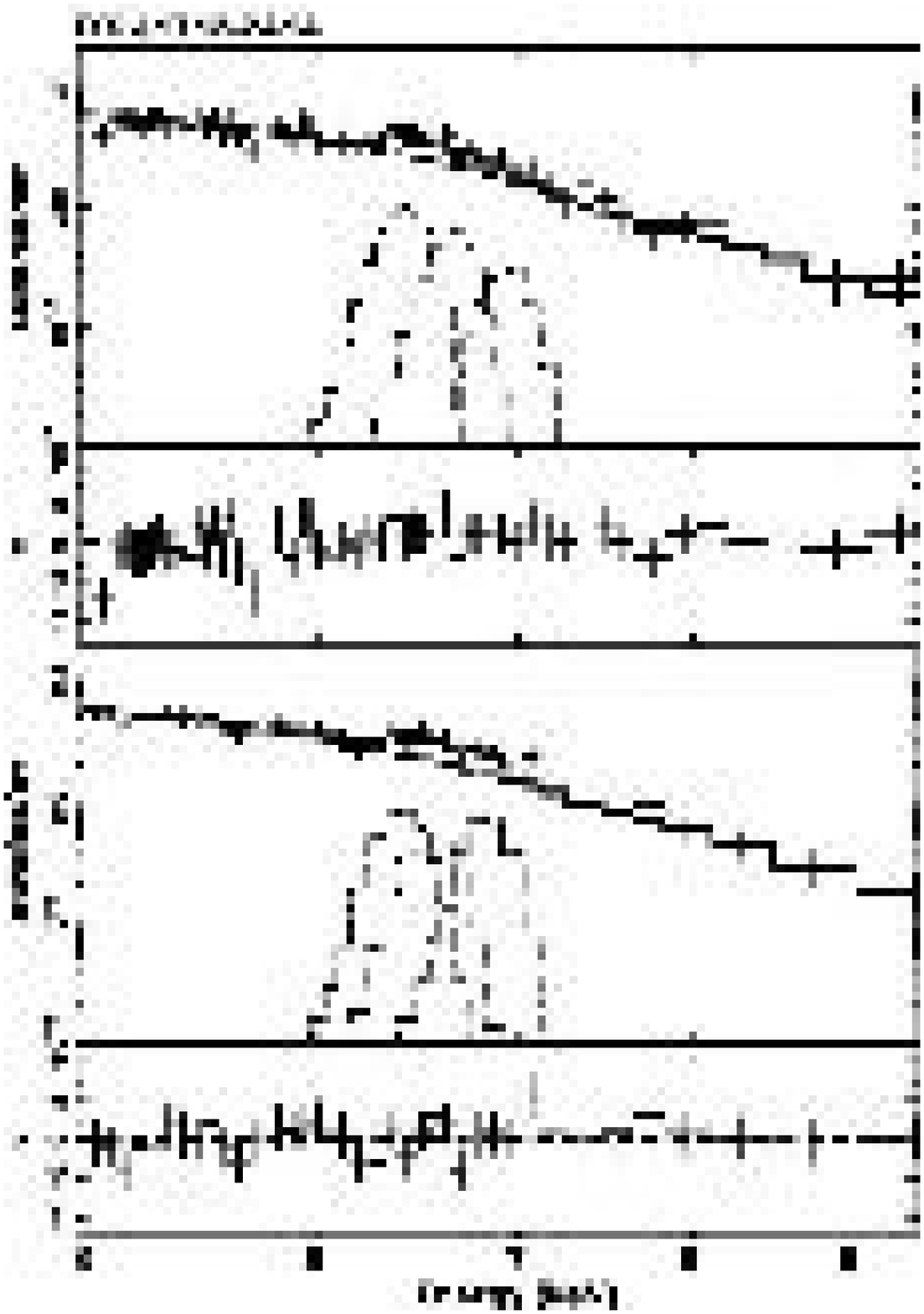}
\FigureFile(7.5cm,12cm){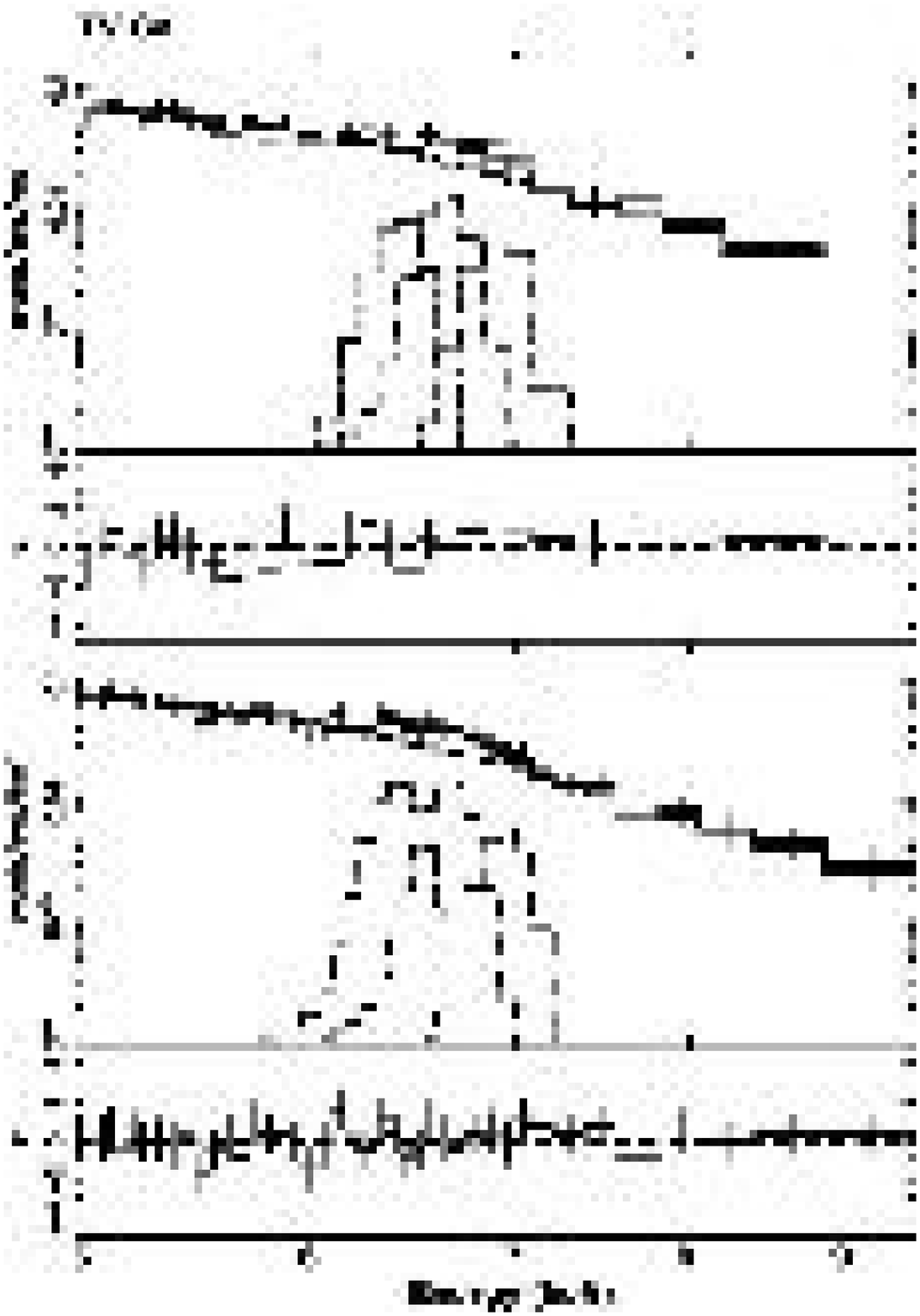}
\FigureFile(7.5cm,12cm){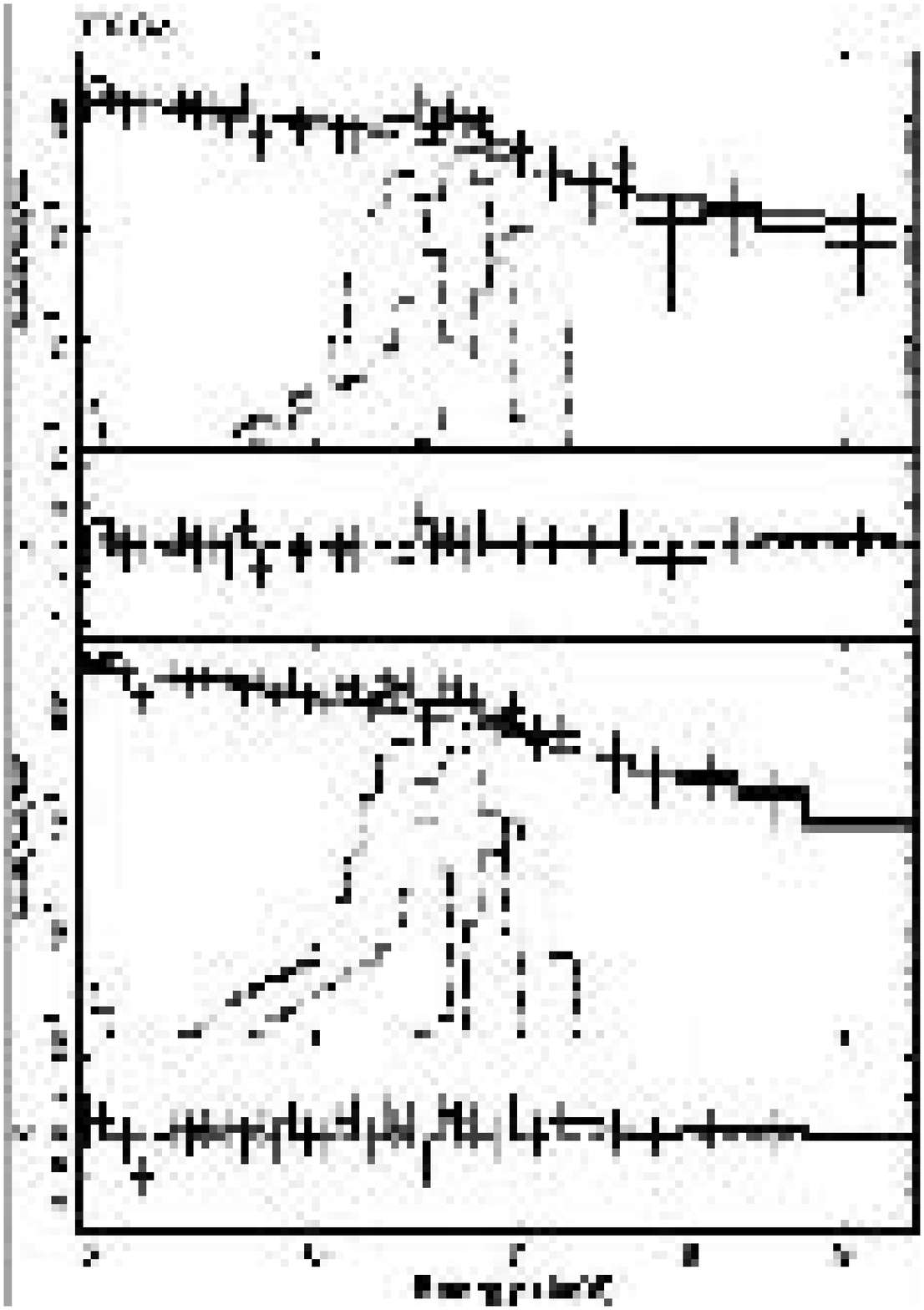}
\end{center}
\centerline{{\bf Figure \ref{fig:intermediate_polars_spec}} Continued.}
\end{figure*}

\begin{figure*}[hbt]
\begin{center}
\FigureFile(7.5cm,12cm){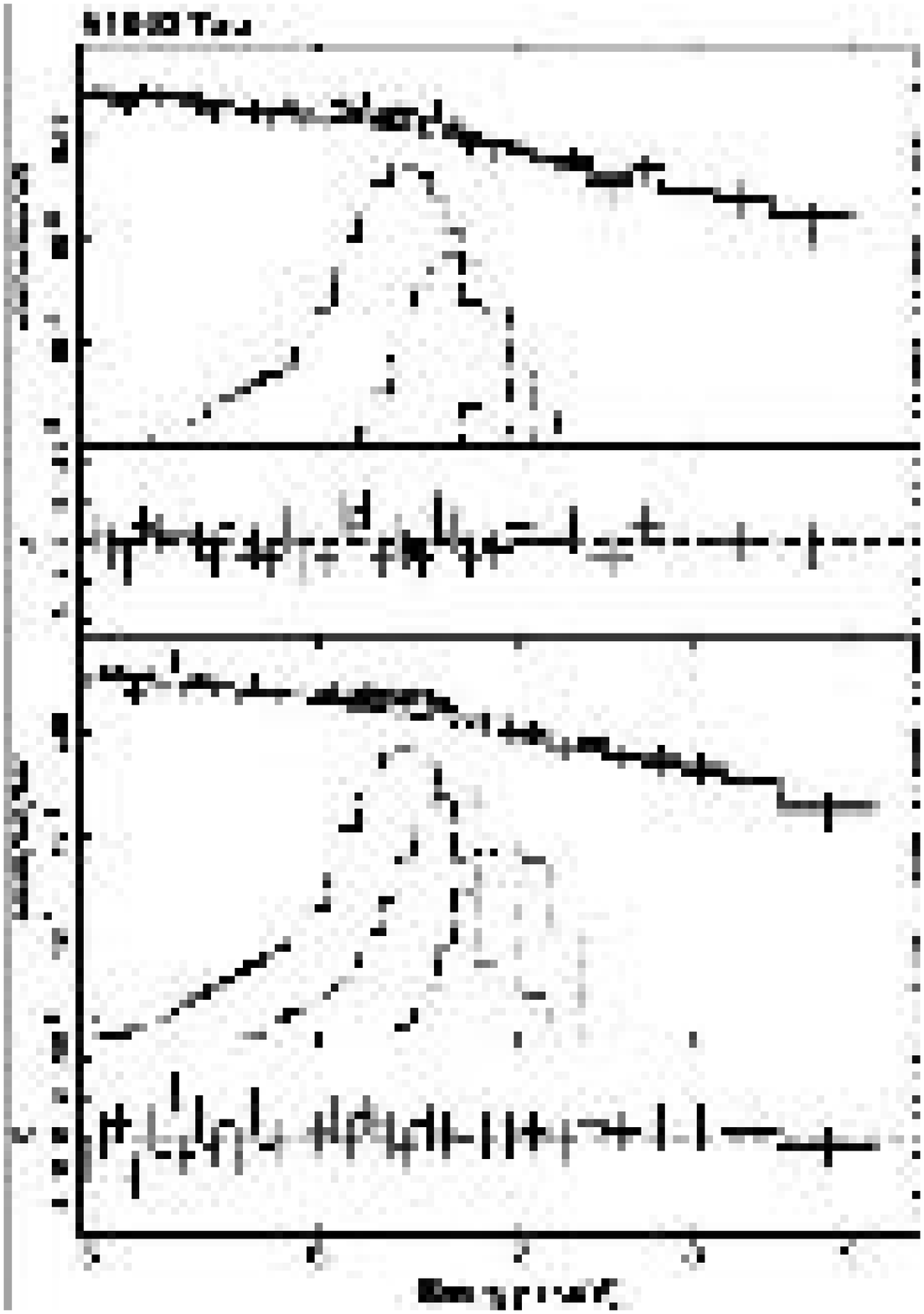}
\FigureFile(7.5cm,12cm){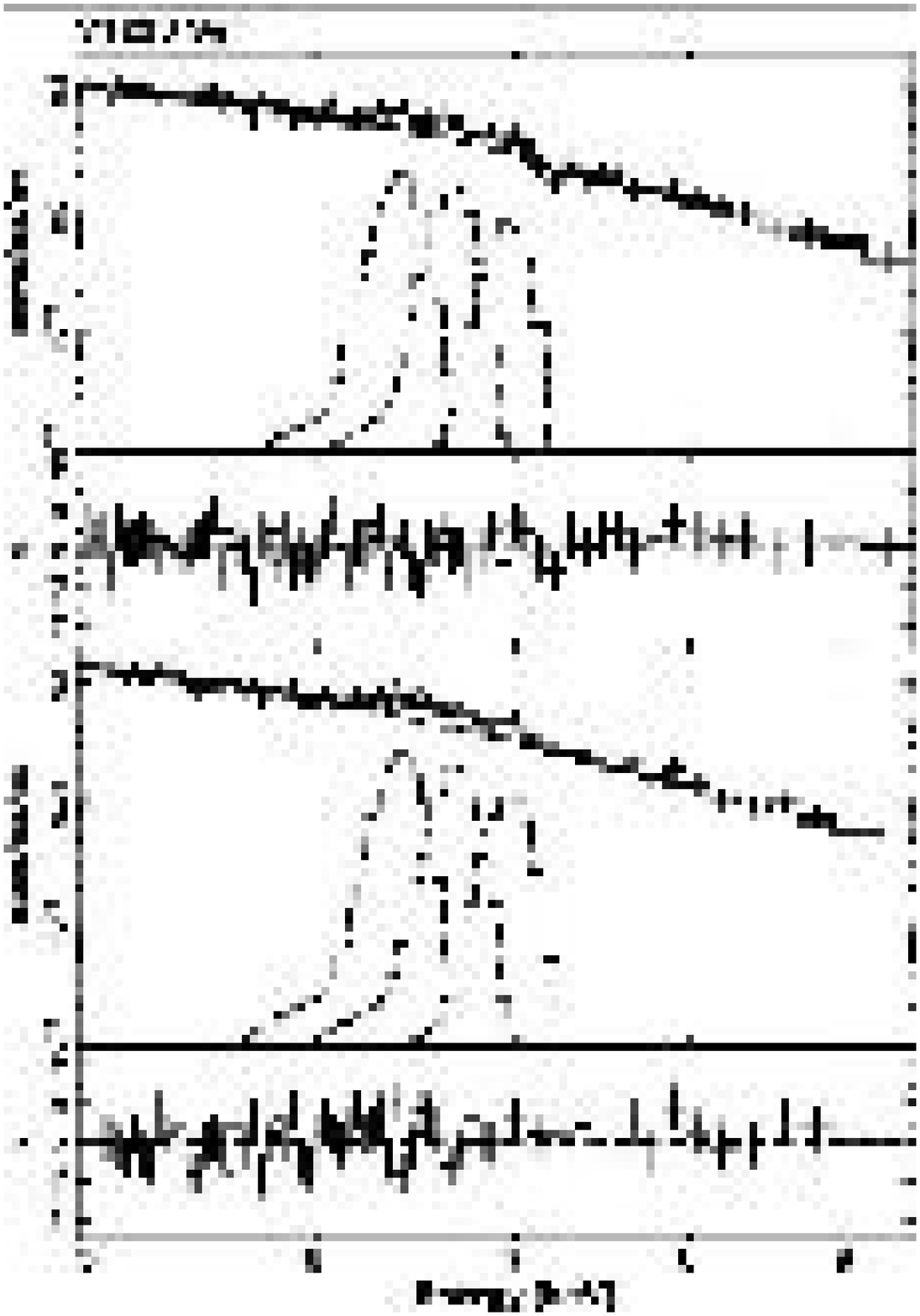}
\FigureFile(7.5cm,12cm){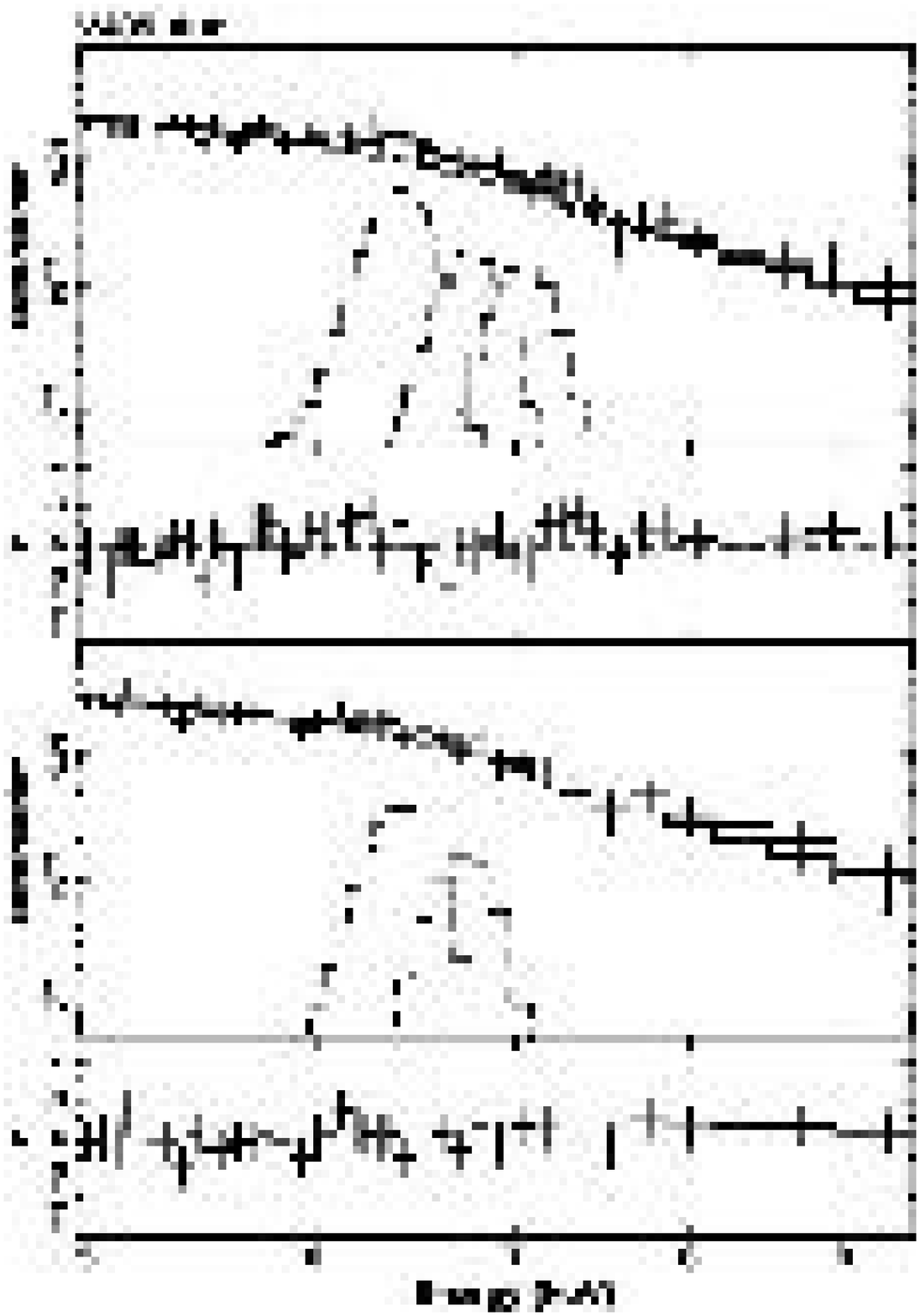}
\FigureFile(7.5cm,12cm){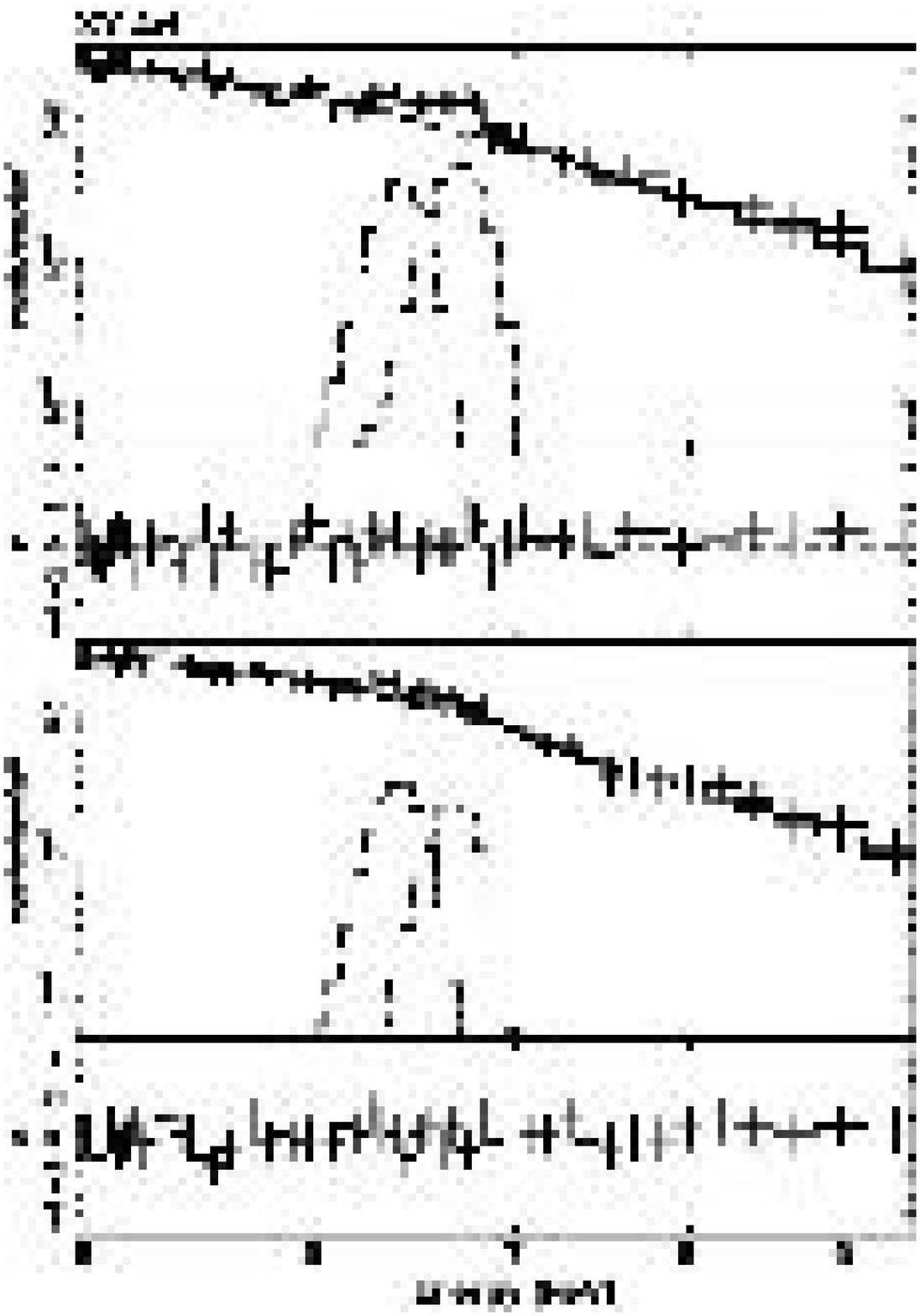}
\end{center}
\centerline{{\bf Figure \ref{fig:intermediate_polars_spec}} Continued.}
\end{figure*}

\begin{figure*}[hbt]
\begin{center}
\FigureFile(8.0cm,6.0cm){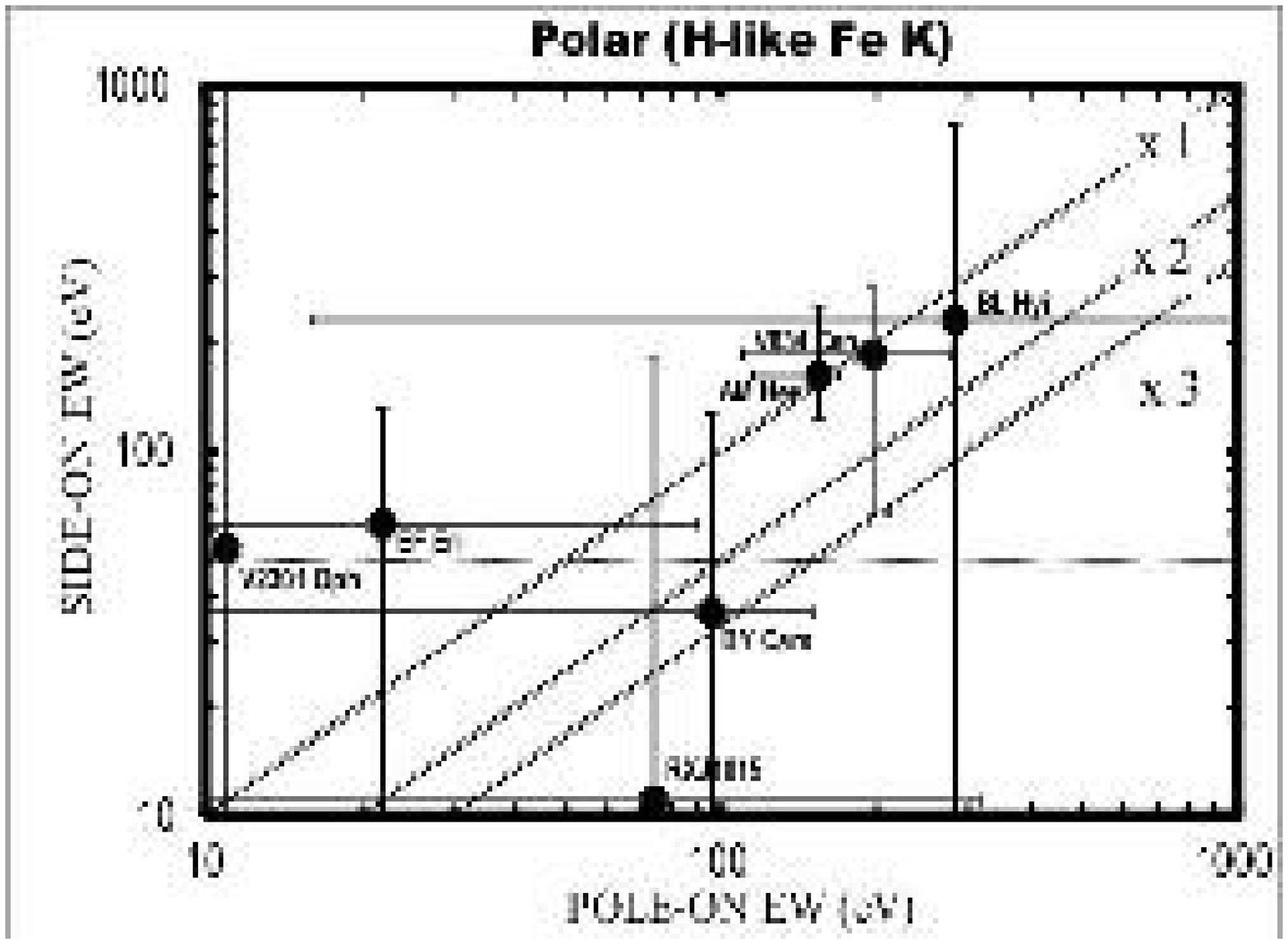}
\FigureFile(8.0cm,6.0cm){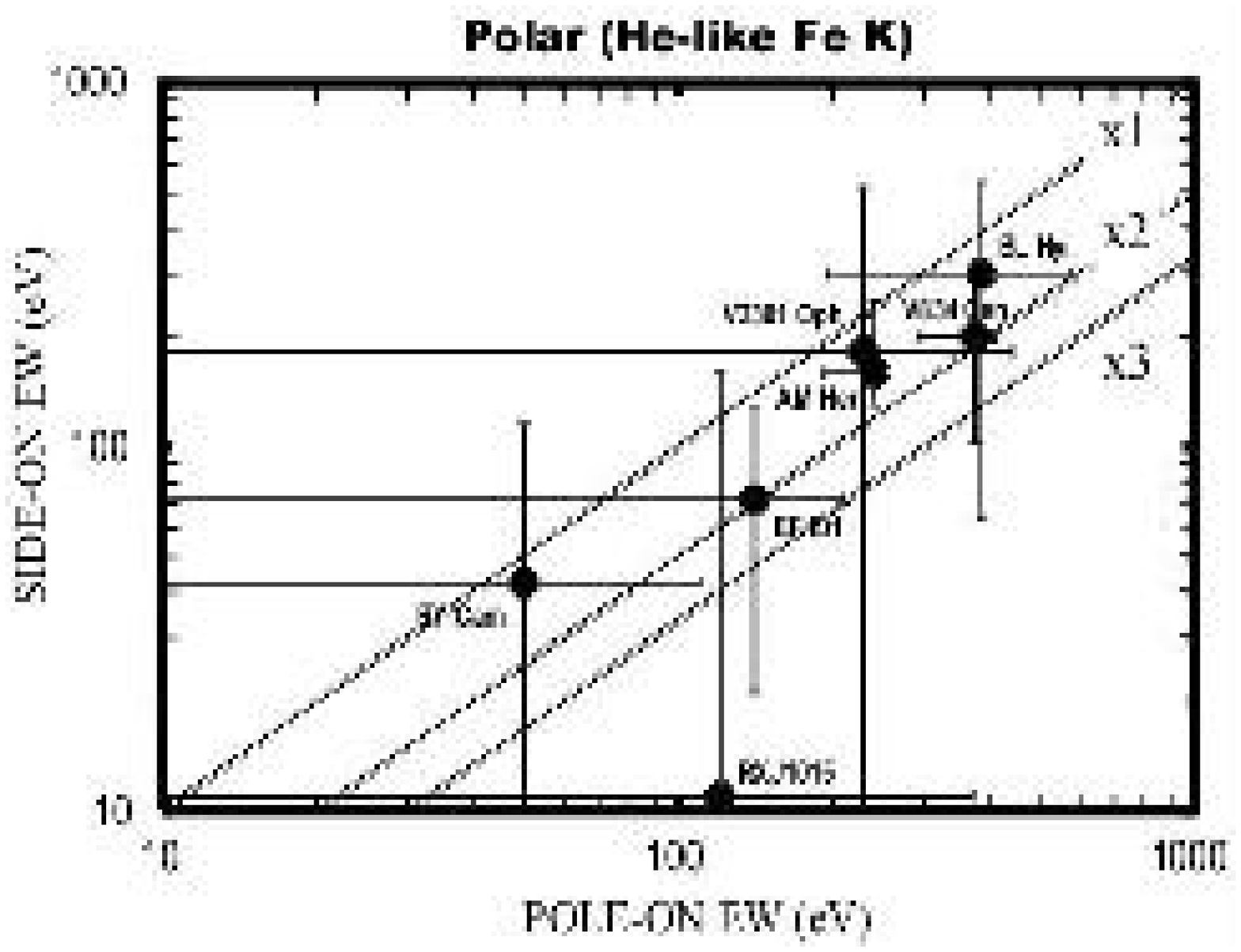}
\FigureFile(8.0cm,6.0cm){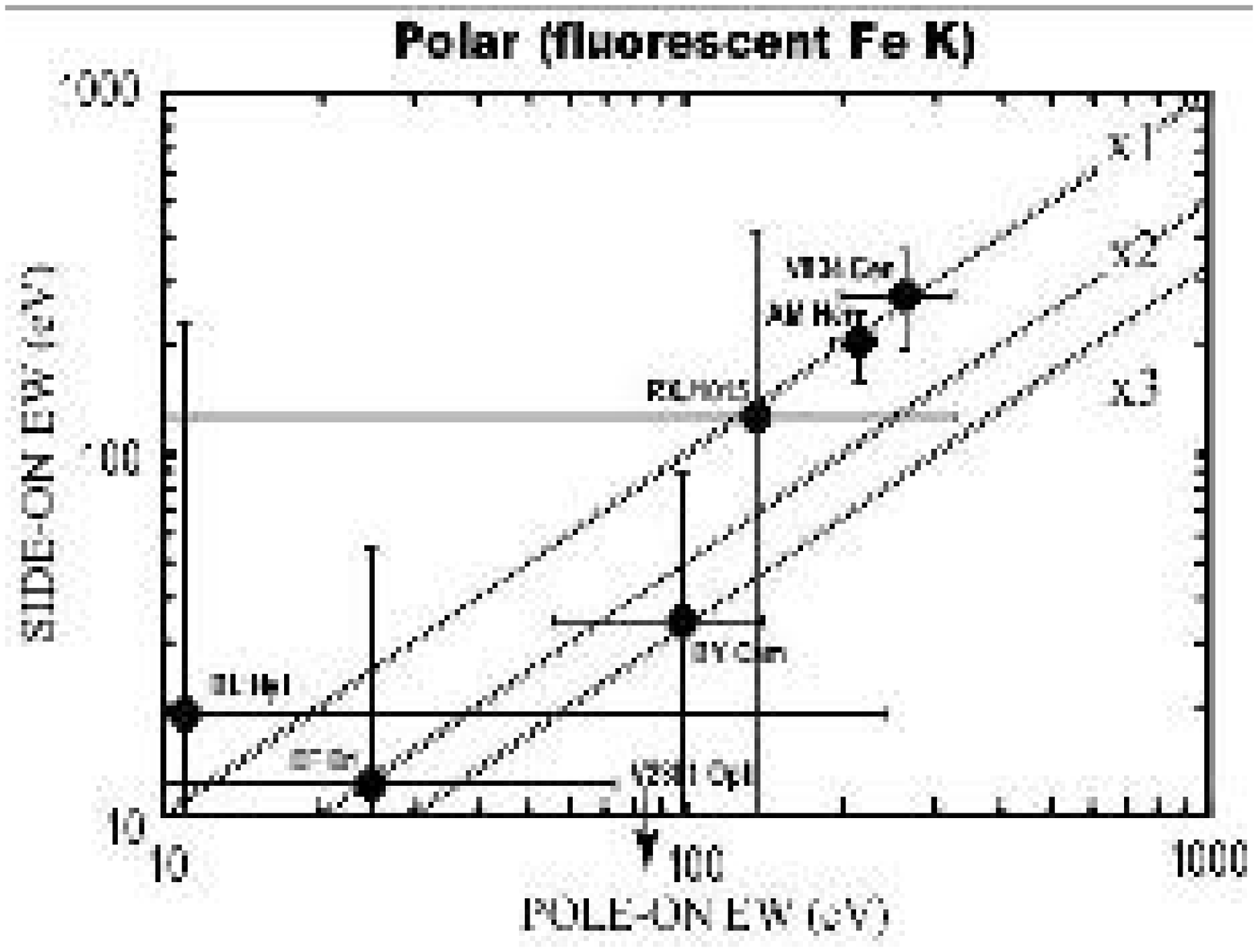}
\end{center}
\caption{The Fe line EWs of the polars measured at the pole-on phase, 
compared with those at the side-on phase.
The top left and top right panels are for H-like and He-like Fe K lines,
respectively. For reference, EWs of fluorescent Fe K lines 
are plotted in the bottom panel.}
\label{fig:polars_ew}
\end{figure*}

\begin{figure*}[hbt]
\begin{center}
\FigureFile(8.0cm,6.0cm){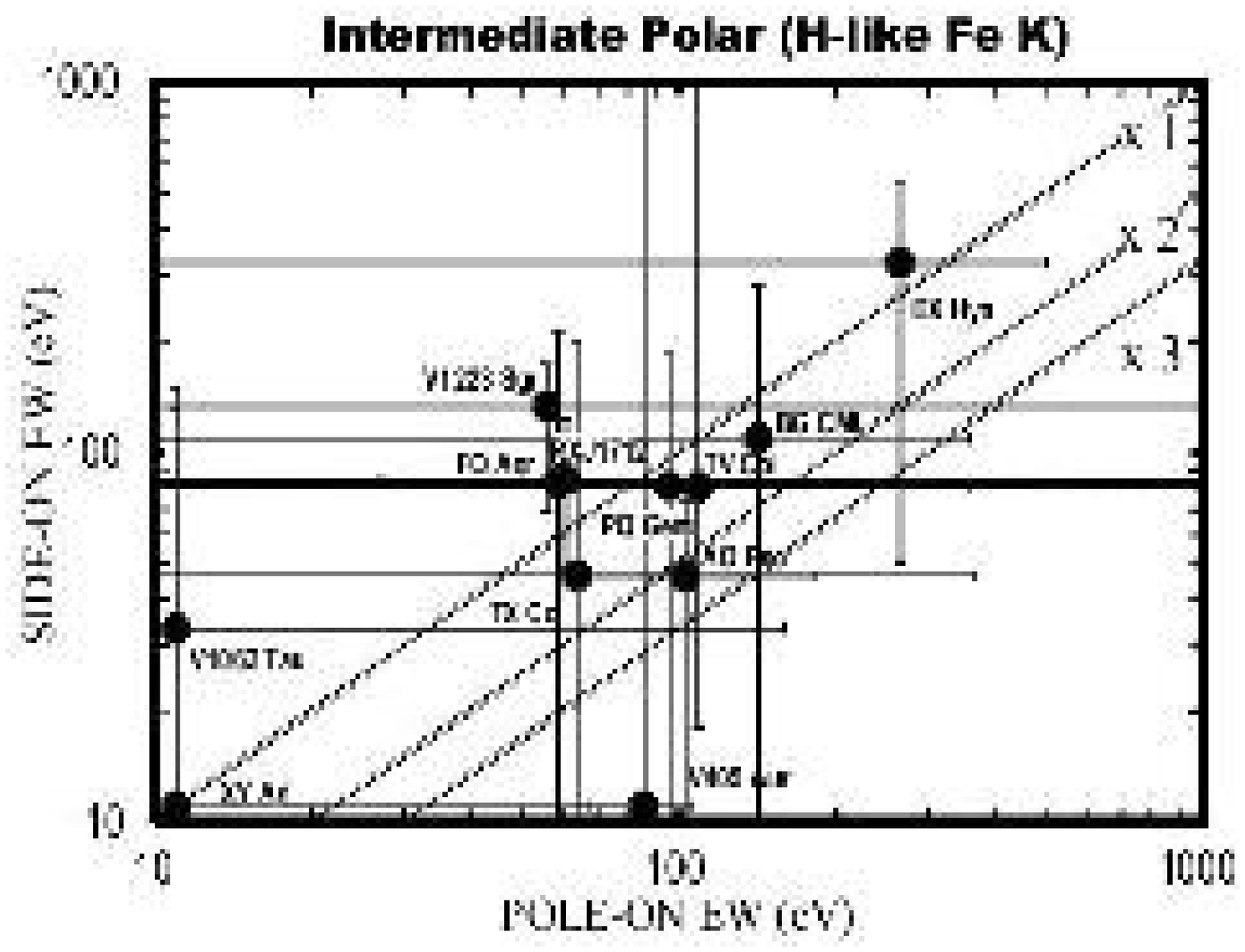}
\FigureFile(8.0cm,6.0cm){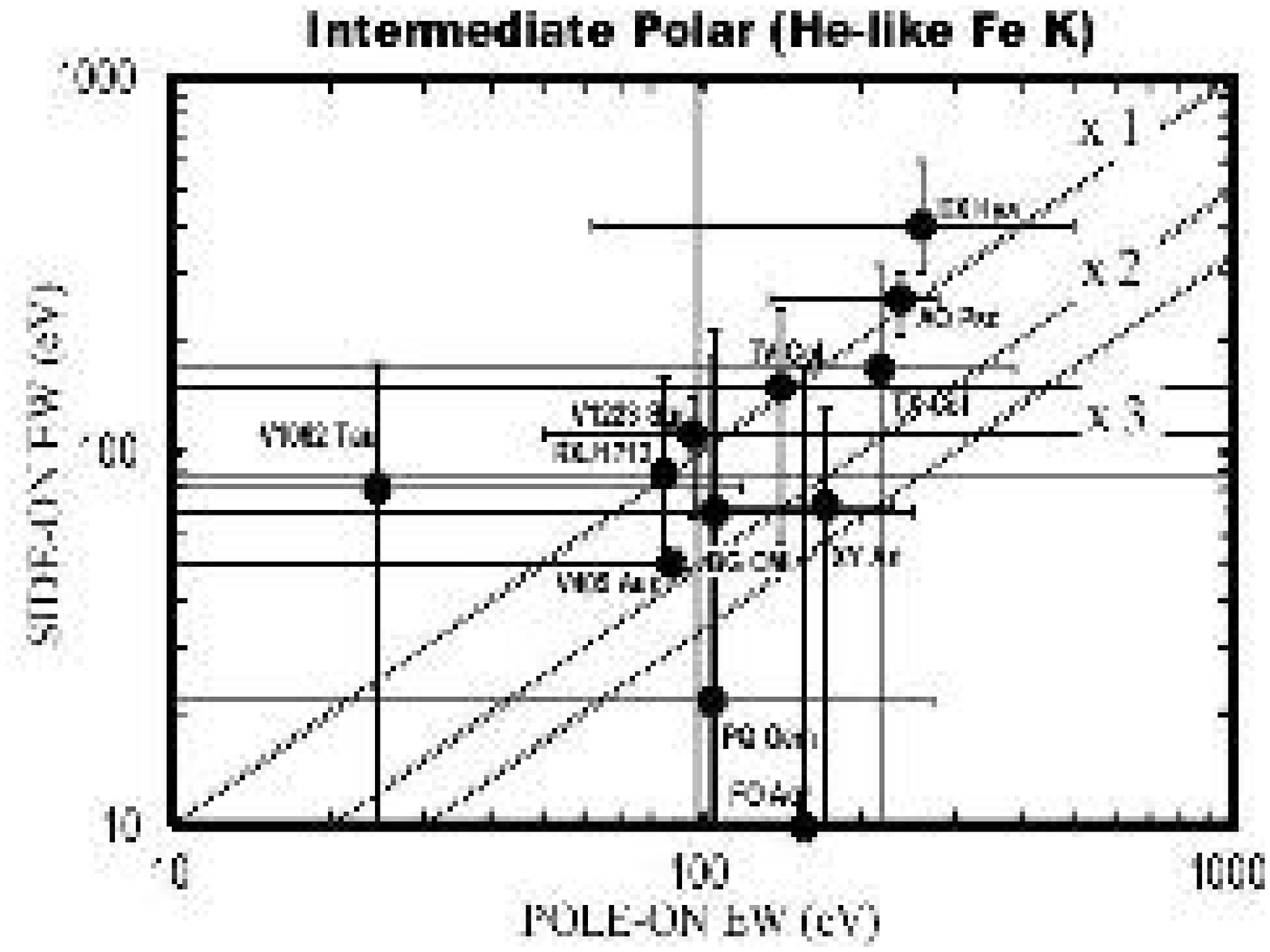}
\FigureFile(8.0cm,6.0cm){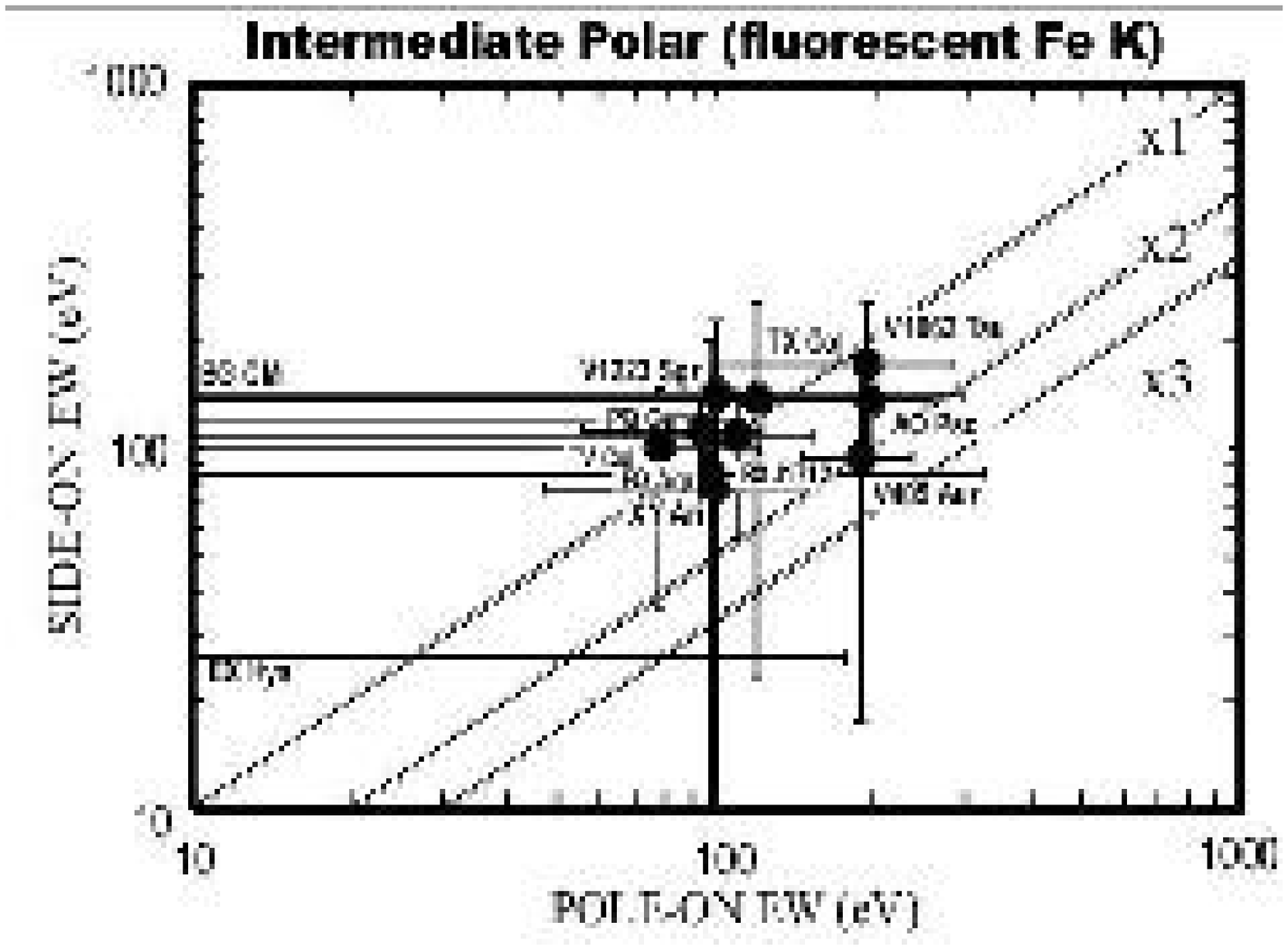}
\end{center}
\caption{The same as figure \ref{fig:polars_ew}, but for the IPs.}
\label{fig:intermediate_polars_ew}
\end{figure*}

\begin{figure*}[hbt]
\begin{center}
\FigureFile(8.0cm,6.0cm){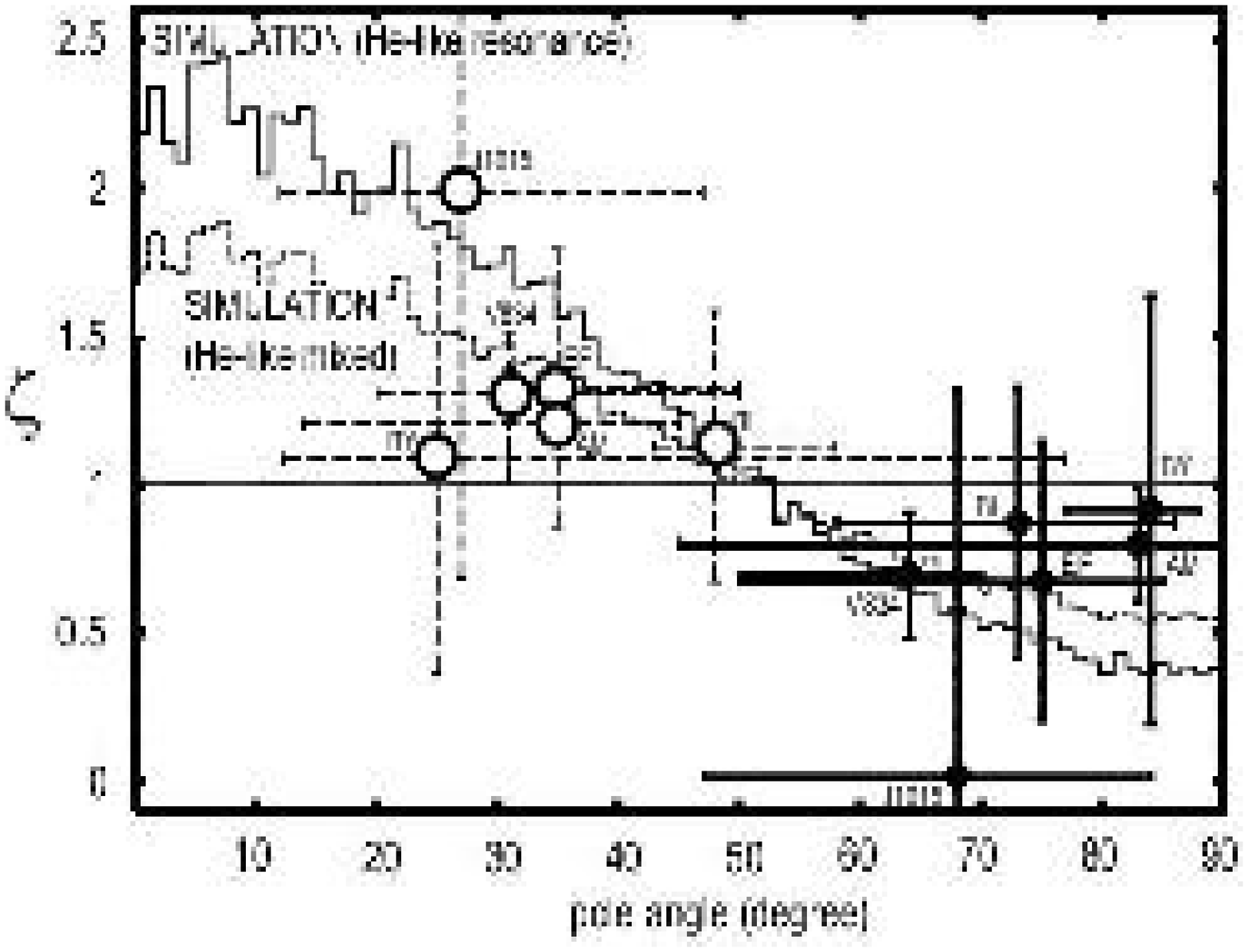}
\FigureFile(8.0cm,6.0cm){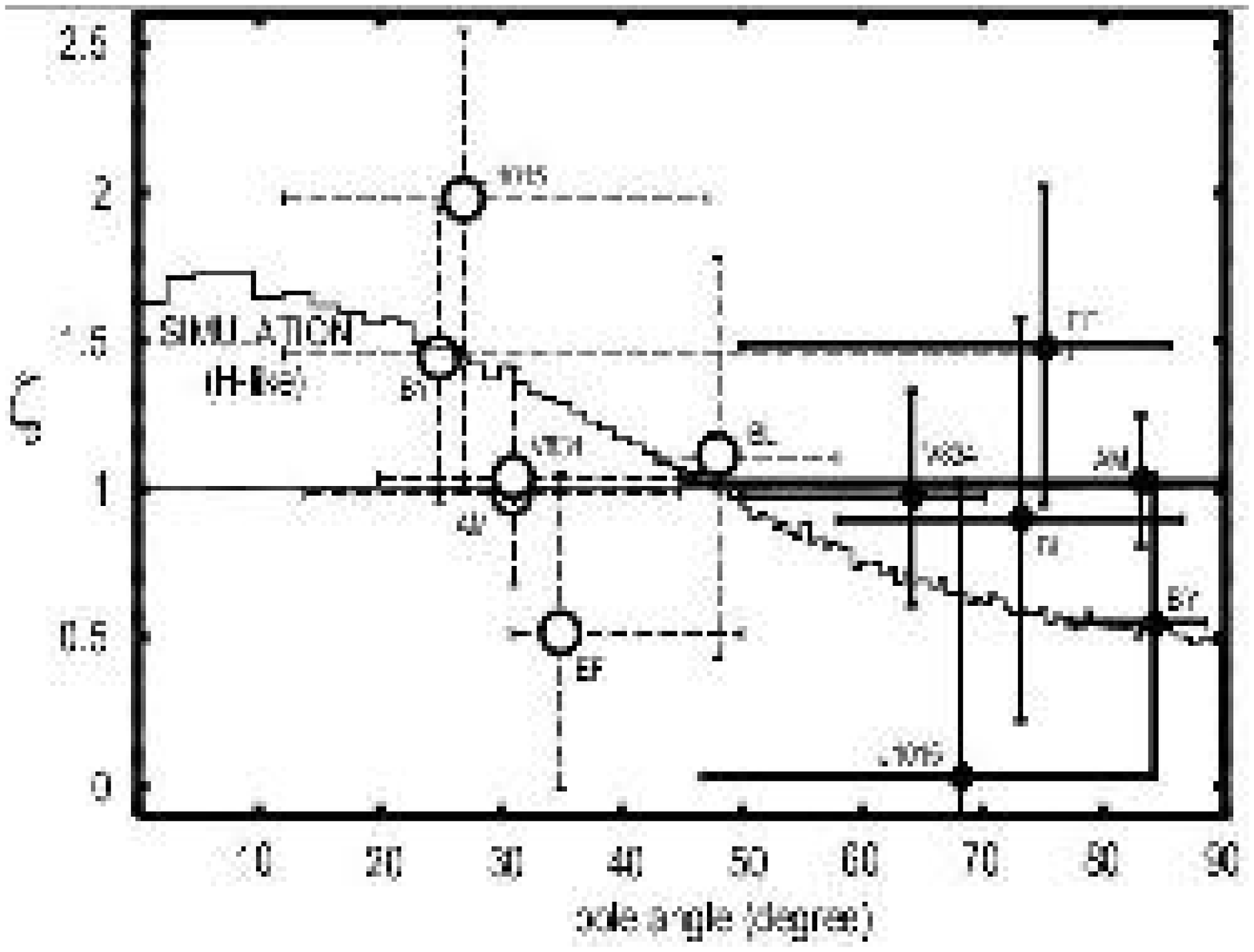}
\end{center}
\caption{The EWs of helium-like (left) and H-like (right) Fe lines 
of the 7 polars in their pole-on and side-on phases, 
relative to the phase-averaged value. The data of V2301 Oph are not plotted, 
because the geometrical parameters are not available.
They are presented against the pole angle $\theta$,
which is calculated using the geometrical parameters 
in table 4 of Paper I. Each object appears twice,
with the circle for pole-on data
and with the filled diamond for side-on data.
The angular distributions predicted by the Monte Carlo simulation
(Paper I) in the nominal case are also plotted, 
assuming a temperature of 16 keV, the bulk velocity of the plasma 
of 0.9$\times 10^8$ cm s$^{-1}$, the electron density
of 7.7$\times 10^{15}$ cm$^{-3}$ just below the shock front, and
the column radius of 7.0$\times 10^7$ cm.}
\label{fig:polar_theta_xi}
\end{figure*}

\begin{figure*}[hbt]
\begin{center}
\FigureFile(8.0cm,6.0cm){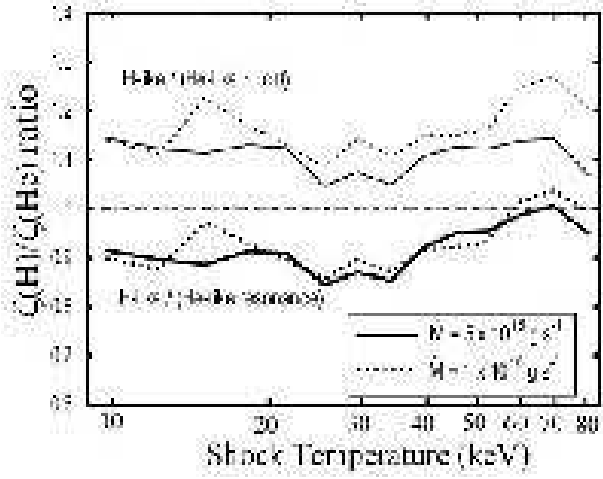}
\end{center}
\caption{Comparison of the anisotropic effects for the H-like and He-like 
Fe K$_\alpha$ lines, based on our Monte-Carlo simulation. 
The ratio of the enhancement $\zeta$ 
at the exact pole-on direction ($\theta = 0$) 
of H-like and He-like Fe K$_\alpha$ lines is
shown as a function of mass-accretion rate 
$5 \times 10^{16}$ g s$^{-1}$ and $1 \times 10^{16}$ g s$^{-1}$.
Calculations were done for the same sets of conditions as Paper I 
(shown in the caption to figure \ref{fig:polar_theta_xi}).}
\label{fig:sim_rslt_he_h_temp}
\end{figure*}

\begin{figure*}[hbt]
\begin{center}
\FigureFile(7.0cm,){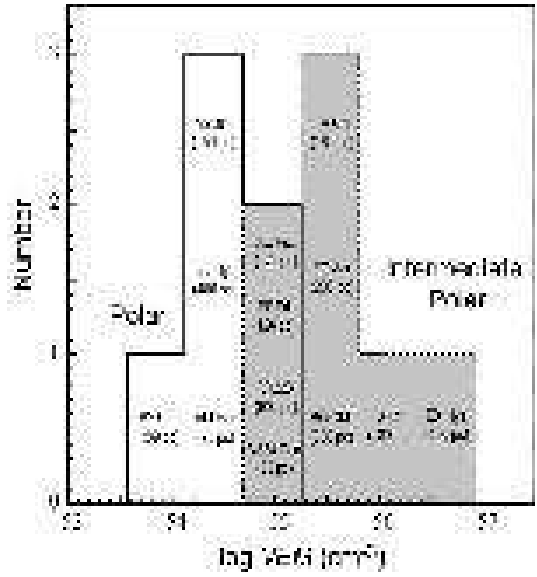}
\end{center}
\caption{The distribution of the measured emission measure of polars
(thick solid histogram) and IPs (dashed lines with shadows) shown in
table \ref{tbl:target_table}. The object names and their distances are
also shown in this figure.}
\label{fig:mcv_emission_measure}
\end{figure*}

\end{document}